\pdfoutput=1
\documentclass[lettersize,journal]{IEEEtran}
\IEEEoverridecommandlockouts
\usepackage{epstopdf}
\usepackage{cite}
\usepackage{amsmath,amssymb,amsfonts,amsthm,enumitem}
\usepackage{graphicx}
\usepackage{textcomp}
\usepackage{float}
\usepackage{amsmath}

\usepackage{placeins}
\usepackage{multirow}
\usepackage{algorithmic}
\usepackage{dblfloatfix}
\usepackage{pifont}

\usepackage[table,xcdraw]{xcolor}
\usepackage[normalem]{ulem}
\useunder{\uline}{\ul}{}
\usepackage{booktabs}
\usepackage[none]{hyphenat}
\usepackage[ruled,vlined]{algorithm2e}

\usepackage{adjustbox}
\usepackage{enumitem}
\usepackage[english]{babel}
\usepackage[utf8]{inputenc}
\usepackage{comment}
\usepackage{subcaption}
\usepackage{array}
\usepackage{url}
\usepackage{verbatim}

\begin{document}
	
\title{Exposing Reliability Degradation and Mitigation in Approximate DNNs under Permanent Faults}

\author{Ayesha Siddique and Khaza Anuarul Hoque~\IEEEmembership{Senior Member,~IEEE} 
\IEEEcompsocitemizethanks{
\IEEEcompsocthanksitem Ayesha Siddique and Khaza Anuraul Hoque are with Department of Electrical Engineering and Computer Science, University of Missouri, Columbia, MO, USA \protect\\
E-mail: \{ayesha.siddique@mail.missouri.edu, hoquek@missouri.edu\}}
\thanks{Manuscript received xxxx, xx}}

\markboth{IEEE Transactions on Very Large Scale Integration (VLSI) Systems, ~Vol.x, No.~x, xxxx}%
{Shell \MakeLowercase{-\textit{A. Siddique and K.A. Hoque}}: Exposing Reliability Degradation and Mitigation in Approximate Deep Neural Networks under Permanent Faults}


\maketitle
\begin{abstract}
Approximate computing is known for enhancing deep neural network accelerators' energy efficiency by introducing inexactness with a tolerable accuracy loss. However, small accuracy variations may increase the sensitivity of these accelerators towards undesired subtle disturbances, such as permanent faults. The impact of permanent faults in accurate deep neural network (AccDNN) accelerators has been thoroughly investigated in the literature. Conversely, the impact of permanent faults and their mitigation in approximate DNN (AxDNN) accelerators is vastly under-explored. Towards this, we first present an extensive fault resilience analysis of approximate multi-layer perceptrons (MLPs) and convolutional neural networks (CNNs) using the state-of-the-art Evoapprox8b multipliers in GPU and TPU accelerators. Then, we propose a novel fault mitigation method, i.e., fault-aware retuning of weights (Fal-reTune). Fal-reTune retunes the weights using a weight mapping function in the presence of faults for improved classification accuracy. To evaluate the fault resilience and the effectiveness of our proposed mitigation method, we used the most widely used MNIST, Fashion-MNIST, and CIFAR10 datasets. Our results demonstrate that the permanent faults exacerbate the accuracy loss in AxDNNs compared to the AccDNN accelerators. For instance, a permanent fault in AxDNNs can lead to 56\% accuracy loss, whereas the same faulty bit can lead to only 4\% accuracy loss in AccDNN accelerators. We empirically show that our proposed Fal-reTune mitigation method improves the performance of AxDNNs up to 98\%, even with fault rates of up to 50\%. Furthermore, we observe that the fault resilience in AxDNNs is orthogonal to their energy efficiency.

\end{abstract}
	

\maketitle 
\section{Introduction}
\label{sec:introduction}

\IEEEPARstart{D}{eep} learning algorithms have the potential to add high levels of autonomy to industry 4.0 systems due to their ability to process and classify big-data in different applications, such as smart healthcare, automotive, smart factories, etc. Such algorithms are primarily based on deep neural networks, which have been progressing through different performance requirements lately. For example, a variety of CNN models, such as AlexNet \cite{krizhevsky2012imagenet}, VGG \cite{simonyan2014very}, etc., addressing the computational and memory challenges at multiple hardware platforms have been developed \cite{capra2020hardware}. The most widely used hardware platforms include Graphic processing units (GPU) and Tensor processing units (TPU) \cite{jouppi2017datacenter} that inherit massive data parallelism. Though the robustness of these hardware platforms is very appealing to modern DNNs, their enormous computational power comes at the cost of high energy consumption magnified by the cooling loads \cite{zervakis2022thermal, malawade2021sage}. Consequently, their deployment as DNN accelerators in smart nanoscale applications becomes very challenging \cite{hsu2021gptpu} e.g., complex DNN analytics in resource-constrained edge devices, where safety and energy are critical considerations, can lead to an unexpected energy outage which can jeopardize human lives. This problem can be solved by leveraging approximate computing \cite{zervakis2021approximate, armeniakos2022hardware} -- an inexact computing method that exploits the inherent error resilience of onboard applications for energy efficiency in DNN accelerators. However, approximate hardware acceleration is deemed to be inherently less reliable \cite{rodrigues2019assessing}. 

Approximate computing-based deep neural network (AxDNN) accelerators are designed by incorporating inexact arithmetic units \cite{hanif2019cann} \cite{riaz2020caxcnn}, computation skipping \cite{ganapathy2017dyvedeep}, memory skipping \cite{song2019approximate}, etc. Their fabrication at the nanoscale follows a sophisticated manufacturing process whose imperfections may result in manufacturing defects, such as process variations and permanent faults (stuck-at faults) \cite{armeniakos2022hardware}. As discussed in this paper, the permanent faults affect the compute units of AxDNN accelerators in every execution cycle and their presence as unmasked faults leads to serious failures in the whole system. Indeed, their impact is stronger in AxDNN accelerators due to self-error-inducing approximate computations compared to the accurate design alternatives, i.e., accurate deep neural network (AccDNN) accelerators \cite{siddique2021exploring}. However, the ratio of fault resilience in AxDNNs to that of AccDNNs depends on data precision, location and type of permanent faults, size of accelerators, degree of approximations, activation functions, neural network topology, and accelerators. With so many variables to consider, analyzing the reliability of AxDNNs against permanent faults becomes very challenging. 

The permanent faults are usually detected using post-fabrication testing for discarding the faulty manufactured chips. However, if a high number of manufactured chips are faulty, discarding them reduces the yield to a large extent. A potential solution is employing redundant executions (re-execution) to ensure correct outputs \cite{vidya2022softsnn}, but retraining AxDNNs increases the
overall runtime \cite{venkataramani2014axnn}. For example, training an AxDNN with 50 layers having arbitrary approximate components takes 89 days \cite{mrazek2016design}. The reason is that the simulation of approximate arithmetic components does not usually scale well and their training is thus slowed down by one order of magnitude for CPU and in three orders for GPU \cite{mrazek2019alwann}. Thus, in the current resource-constrained nanoscale hardware paradigm, where approximate arithmetic units are vastly explored for energy efficiency, it is imperative to not only extensively analyze the reliability of AxDNNs but also, maximize their yield with an efficient and fault-tolerant strategy.

Investigating and mitigating the reliability issues in AxDNNs is an interesting research problem. However, the most recent works' main concentration is limited to AccDNNs. Recently, Kundu et al. elucidated the impact of masking and non-masking of permanent faults in small-scaled Google-TPU-like systolic array-based  DNN accelerators with their formal guarantees \cite{kundu2020high}. In another work, Santos et al. injected the permanent faults in the register files of GPU to demonstrate their effect on reduced precision AccDNNs \cite{dos2018analyzing}. Guerrero-Balaguera et al. analyzed their impact in both register files and the functional units of GPU running AccDNNs \cite{guerrero2022reliability}. Condia et al. studied the effects of faults in critical and user-hidden modules (such as the Warp Scheduler and the Pipeline Registers) for the convolution computations in AccDNNs over GPU \cite{condia2021combining}. Very recently, our \textit{previous work} in \cite{siddique2021exploring} explored the fault resilience of AxDNNs with their energy trade-offs running on a 256x256 approximate systolic array-based DNN accelerator that is analogous to Google TPU. However, the analysis presented in \cite{siddique2021exploring} has several limitations: (i) the analysis is limited to simple approximate multi-layer perceptrons, (ii) only layer-wise permanent faults on TPU-based AxDNNs are analyzed, and non-layer-wise analysis is missing, (iii) only systolic-array based architecture is analyzed and the impact of permanent faults on GPU-based accelerators is ignored, and (iv) the analysis does not provide any insights on the mitigation of the permanent faults. It is worth recalling that retraining AxDNNs is a complicated problem; hence, the state-of-the-art retraining-based fault mitigation methods \cite{zhang2018analyzing, zhang2019fault, abdullah2020salvagednn} for AccDNNs cannot be directly applied to AxDNNs. In summary, while several efforts to analyze and mitigate the reliability issues in AccDNNs on TPU and GPU-based accelerators have been made, there is a considerable research gap in analyzing and mitigating the impact of permanent faults in AxDNNs mapped on TPU and GPU-based accelerators.



Motivated by this, in this paper, we significantly extend our previous work in \cite{siddique2021exploring} with the following three novel contributions in the context of AxDNNs: (i) we present an extensive fault resilience analysis using a variety of approximate convolutional neural networks (CNNs) and multi-layer perceptrons (MLPs) for TPU and GPU accelerators. Specifically, we analyze the impact of stuck-at faults on approximate Lenet-5, VGG-11, AlexNet, and two multi-layer perceptrons with MNIST, Fashion-MNIST, and CIFAR-10 datasets. Our fault resilience (layer-wise and non-layer-wise) analysis includes different fault types, fault-bit positions, fault layers, activation functions, layer widths, and degree of approximation errors, (ii) we explore the trade-offs between fault resilience and energy consumption in these approximate networks, (iii) we also propose a novel fault mitigation method i.e., \uline{\textbf{fa}}u\uline{\textbf{l}}t-aware \uline{\textbf{retun}}ing of w\uline{\textbf{e}}ights (Fal-reTune). Fal-reTune first performs a fault-aware mapping of AxDNNs on the underlying hardware and then, retunes the weights using a weight mapping function for approximate arithmetic components. Our results show that the faults exacerbate the accuracy loss with approximate computing and fault resilience of AxDNNs varies from output to input layer following the type of activation functions and the amount of approximation error. The higher the approximation error, the higher the fault's tendency to disrupt the output quality. We empirically show that a permanent fault in an approximate multi-layer perceptron can lead up to 56\% accuracy loss. In contrast, the same fault in the same position can lead to only a 4\% accuracy loss in its accurate counterpart. Interestingly, our proposed Fal-reTune mitigation efficiently improves the classification accuracy of the approximate MLPs and CNNs by up to 98\% even with fault rates of up to 50\%. Also, we observe that the fault resilience is orthogonal to the energy consumption in approximate networks.

The remainder of this paper is structured as follows: Section \ref{sec:relatedworks} discusses the state-of-the-art works analyzing the reliability of AccDNNs and AxDNNs. Section \ref{sec:prelim} provides the preliminary information about DNNs, DNN hardware accelerators, and approximate multipliers. Section \ref{sec:methodology} and Section \ref{sec:mitigation} present our evaluation methodology and the proposed Fal-reTune mitigation. Section \ref{sec:results} the results obtained from our extensive fault resilience analysis and Fal-reTune fault mitigation in AxDNNs. Finally, Section \ref{sec:conclusion} concludes the paper.

\section{Related Works}
\label{sec:relatedworks}

The emerging demand for highly reliable neural networks in safety-critical applications has encouraged researchers to deeply analyze the impact of faults on DNNs. The faults can be permanent or transient in nature \cite{sabih2021fault}. The permanent faults affect every execution cycle of DNN accelerators and hence, degrade their performance more significantly than occasional transient faults \cite{kundu2020high}. To this end, the state-of-the-art works present different analysis methods and frameworks for extensively analyzing the impact of permanent faults on AccDNNs \cite{shafique2020robust}. Earlier, Reagen et al. presented a lightweight framework for empirical fault resilience analysis of AccDNNs towards permanent faults, which arise due to process variation or flash lifetime wear problems \cite{reagen2018ares}. The available fault locations are limited to the memory domain and include weights, activities, and hidden states. In another work, Rusopo et al. proposed a permanent fault injection environment built on a darknet open-source framework with different precisions \cite{ruospo2020evaluating}. Several bit-width schemes and data types, i.e., floating-point and fixed-point, are adopted for extensive fault resilience analysis of accurate CNN models \cite{ruospo2020evaluating}. Hong et al. assessed the defect tolerance capability of accurate feed-forward neural networks by injecting permanent faults at the transistor level \cite{temam2012defect}. 

 \begin{figure}[!b]
 	\centering
 	\vspace{-0.05in}
 	\includegraphics[width=0.99\linewidth]{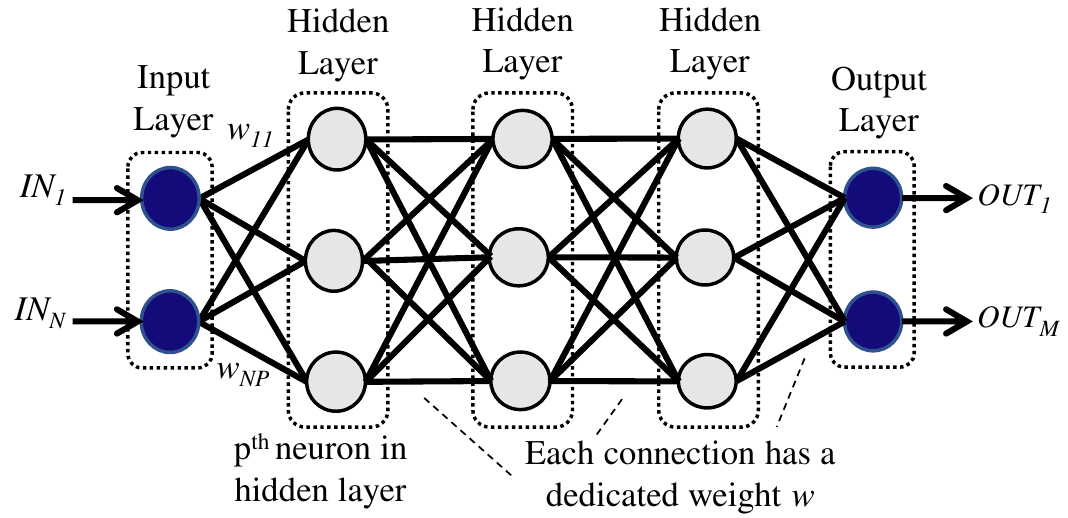}
 	\caption{Fully connected deep neural network}
 	\label{fig:dnn}
 	\vspace {-0.01in}
 \end{figure}

The state-of-the-art works also explore the impact of permanent faults on AccDNN hardware acceleration. Kundu and Hanif et al. elucidated the impact of permanent faults on accurate DNNs running on Google TPU \cite{zhang2018analyzing} \cite{abdullah2020salvagednn}. Zhang et al. studied this impact through formal modeling of stuck-at faults in accurate feed-forward neural networks \cite{kundu2020high}. Other works focused on faults in AccDNNs running on GPU accelerators. Guerrero-Balaguera et al. proposed a framework, resorting to a binary instrumentation tool to perform fault injection campaigns in accurate CNN models, targeting different components inside the GPU, such as the register files and the functional units \cite{guerrero2022reliability}. Tsai et al. presented a tool named NVBitFI, which offers a convenient way to conduct permanent and transient fault injection campaigns into AccDNNs without having the information about their GPU accelerator \cite{tsai2021nvbitfi}. Sio et al. analyzed the reliability of accurate CNN models on programmable hardware of hybrid platforms \cite{de2022firenn}. For mitigating the impact of permanent faults in hardware accelerators, Zhang et al. suggested using fault-aware mapping and pruning of the AccDNN parameters in the underlying hardware such as TPU and then, retrain them for an improved classification accuracy \cite{zhang2018analyzing}. Hanif et al. proposed to enhance this mitigation method through the rearrangement of the AccDNN parameters in the retraining phase at the software level based on the fault locations in the hardware \cite{abdullah2020salvagednn}. However, retraining approximate hardware with approximate multipliers is itself challenging \cite{vaverka2020tfapprox}. Furthermore, another work exploited the weight ranges for bypassing the faulty multipliers and MAC units in AccDNNs \cite{zhang2019fault}; however, the applicability of this fault mitigation method to AxDNNs depends also on the approximate training method. In short, most state-of-the-art works are focused on analyzing and mitigating the reliability threats in AccDNNs. However, a considerable research gap exists in investigating and mitigating the fault resilience of approximate CNN models on systolic arrays and GPU-based hardware accelerators. Motivated by this, we extensively investigate and mitigate the impact of stuck-at 0 and 1 faults in AxDNNs.

\section{Background}
\label{sec:prelim}
This section provides a brief overview of the deep neural networks, their hardware accelerators, such as GPU, and TPU, and approximate multipliers for a better understanding of the paper.

\subsection{Deep Neural Networks}

Consider a simple fully connected neural network having a stack of $L$ fully connected layers as shown in Fig. \ref{fig:dnn}. Each layer $l \in [1, L]$ possesses $N$ neurons whose outputs are named as activations $a$. The layer $l$  multiplies the activations from the previous layer $l-1$ with the weight matrix $w$, having dimensions $N x N$ and adds constant biases $b$, having dimension $N x 1$, followed by an activation function $\phi$. Mathematically, this can be expressed as: $y_i^l = \sum$ , where $\phi$ can be ReLU, tanh, sigmoid etc. A most commonly used special case of fully connected layers is convolutional layers in CNN models. Such layers process activations as 3-dimensional tensors, i.e., height $\times$ width $\times$ channels. Also, the weights are processed as 4-dimensional vectors, i.e., height x width x channels x count, where count specifies the number of filters applied to the same input. The output of the convolutional layers possesses the same layout as the input activations; however, the height and width are determined according to the shape of the kernel. In the case of GPU implementation, the matrix multiplications in both convolutional and fully connected layers are processed as a typical tiled general matrix-matrix multiplication (GEMM) while taking advantage of thread parallelism. This eventually speeds up the emulation.

 \begin{figure}[!h]
 	\centering
 	\vspace{-0.01in}
 	\includegraphics[width=1\linewidth]{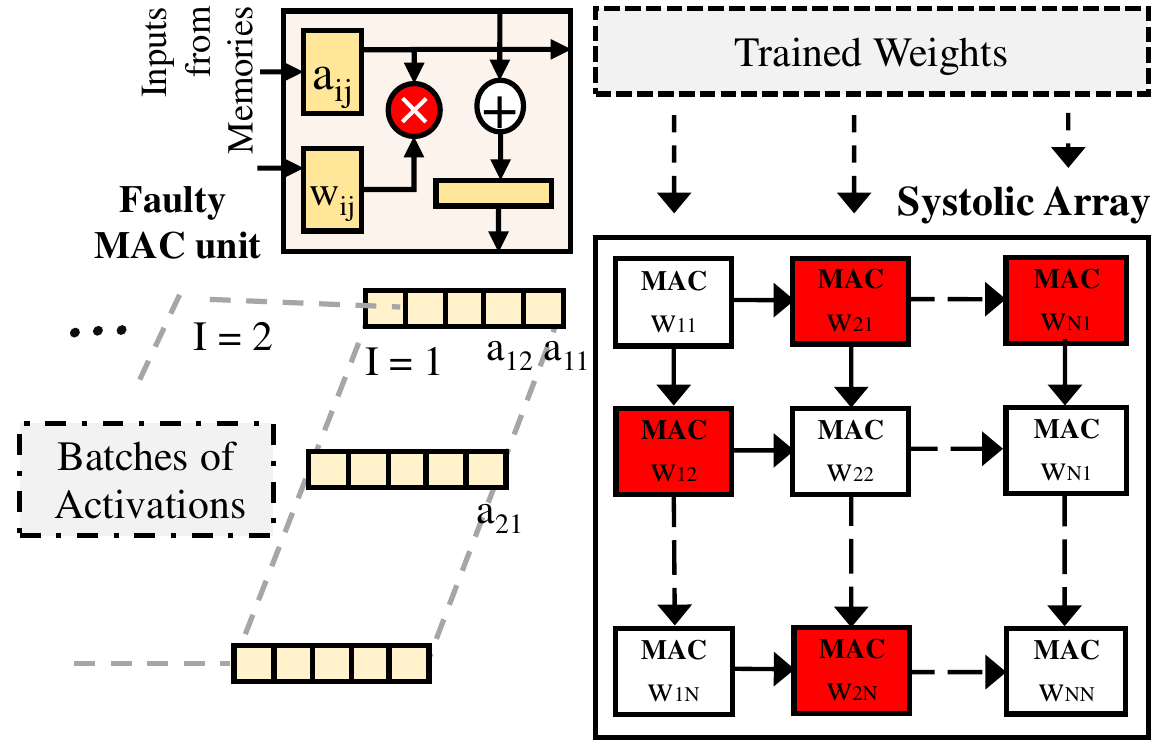}
 	\caption{Systolic array with $N \times N$ grid of multiply and accumulate (MAC) units. The red color in systolic array indicates faulty MAC units with faulty multiplication}
 	\label{fig:systollicarray}
 	\vspace {-0.15in}
 \end{figure}

\begin{figure*}[!t]
	\centering
	\vspace{-0.01in}
	\includegraphics[width=1\linewidth]{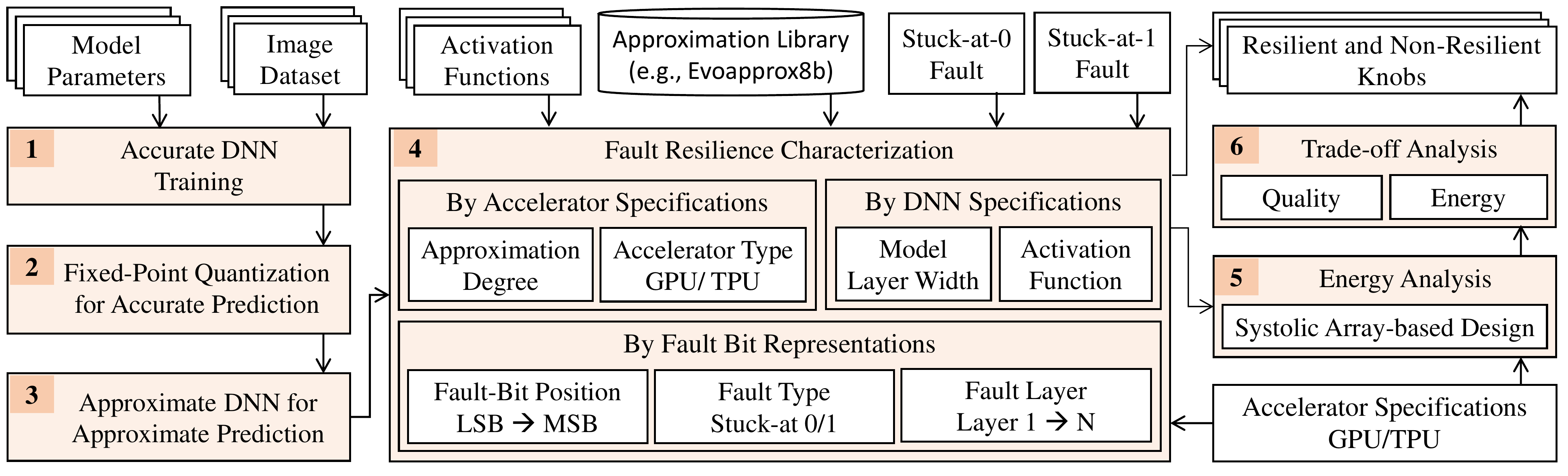}
	\caption{Evaluation methodology for analyzing fault and energy trade-offs in systolic array-based DNN accelerators}
	\label{fig:methodlogy}
	\vspace{-0.01in}
\end{figure*}

\subsection{DNN Hardware Accelerators}
The most commonly used hardware accelerators for DNN hardware acceleration include GPU, and TPU \cite{jouppi2017datacenter}. The TPU employs a two-dimensional systolic array which consists of a $N \times N$ grid of multiply and accumulate (MAC) units as shown in Fig. \ref{fig:systollicarray}. It performs parallel matrix multiplications in fully connected and convolutional layers. To understand the working principle of a TPU, let's consider a fully connected layer with $N$ input and $N$ output neurons and consequently, an $N \times N$ weight matrix loaded into a systolic array. The $w_{i,j}$ in weight matrix is first loaded to the MAC$_{i,j}$ located in $i^{th}$ row and $j^{th}$ column. Then, the activations are loaded into the systolic array to move from the left to the right side periodically. In the first clock cycle, the MAC$_{1,1}$ in the first column of the systolic array computes $w_{1,1}a_1$ and passes the result to the MAC$_{1,2}$ downwards. In the next clock cycle, MAC$_{1,2}$ adds the product $w_{1,2}a_2$ to its input and forwards the result further downwards. After $N$ clock cycles, the MAC$_{1,N}$ finally outputs $w_{1,i}a_i$, which is the first element of weight matrix. Note that the second column also receives and processes the same stream of inputs as the first but is delayed by one clock cycle. This process is replicated along the width of the systolic array. Note that the input matrices larger than the size of the systolic array are processed through the systolic arrays in multiple batches on TPU. On the other hand, GPUs ease the implementation of massive matrix multiplications with multiple threads for robust DNN implementation in the hardware. They employ a typical tiled general matrix-matrix multiplication (GEMM), where the threads load a 2D tile (blocks of filters and channels) from each input matrix into the shared memory. Each thread computes a single output value \cite{vaverka2020tfapprox}. The tiles in the shared memory are quantized to avoid possible shared memory access conflicts. The recent state-of-the-art works focus on introducing tiling mechanism in systolic arrays of TPU \cite{kung2018adaptive, lym2020flexsa}. This paper uses a TPU with a 256 x 256 systolic array similar to Google TPU.

\subsection{Approximate Multipliers}
\label{subsubsec:approxmult}
Approximate multipliers are widely being advocated for energy-efficient computing in error-resilient applications \cite{javed2018approxct}. They incorporate approximate computing techniques, such as approximate counters or compressors \cite{hanif2019cann}, partial products' reduction \cite{zervakis2016design}, and truncation of the carry propagation chain  \cite{ullah2018smapproxlib} etc. for low latency and power or energy consumption. Recently, many opensource libraries for approximate multipliers have been developed e.g., SMApproxLib \cite{ullah2018smapproxlib}, DeMAS \cite{prabakaran2018demas}, Evoapprox8b \cite{mrazek2017evoapproxsb}, lpAClib \cite{rehman2016architectural}, leAp \cite{ebrahimi2020leap}, SIMDive \cite{ebrahimi2020simdive}, etc. Among these, Evoapprox8b is the most extensive library containing different signed and unsigned approximate versions of 28 exact multipliers. These multipliers are ASIC-oriented and built using multi-objective Cartesian genetic programming \cite{ullah2018smapproxlib}. AxDNNs incorporate such approximate multipliers as lookup tables in their models by employing Tensorflow-compatible approximate 2D convolution \cite{vaverka2020tfapprox}. In this paper, we implement AccDNNs and AxDNNs using signed accurate and approximate multipliers, respectively, from Evoapprox8b \cite{ullah2018smapproxlib} library.



\section{Reliability Evaluation Methodology}
\label{sec:methodology}

Our reliability evaluation methodology fulfills three roles: the first, it implements AxDNNs for the hardware accelerators; the second, it characterizes the fault resilience of AxDNNs with fault-bit representations, specifications of models, and accelerators; and the third, it explores the fault-energy trade-offs in AxDNNs. Fig. \ref{fig:methodlogy} presents an overview of our evaluation methodology.

The foremost step of our evaluation methodology is to train AccDNNs with baseline accuracy. For this purpose, we determine the design parameters, e.g., number of hidden layers and number of neurons, learning rate, momentum, etc., through a hit and trial method for different datasets. Then, we quantize the inputs to the most energy-intensive fully connected and convolutional layers in AxDNNs, i.e., approximate MLP and CNN models, respectively. However, this quantization is followed by dequantization, as seen in TensorFlow. In particular, we employ integer 8-bit signed quantization that supports 8-bit signed multiplications. Next, we replace the accurate multipliers in AccDNNs with different approximate multipliers from Evoapprox8b \cite{mrazek2017evoapprox8b} library to design multiple AxDNNs. Typically, the lookup table of an 8-bit multiplier occupies 128 kilobytes only on GPU \cite{vaverka2020tfapprox}. Therefore, we eliminate the need to calculate the output of the approximate multiplier for each input by utilizing the lookup table and speeding up AxDNN simulations on GPU at a very low memory cost. We decompose the layers into a set of tiles to implement AxDNNs with GPU accelerators. Interestingly, such tiles can be mapped to a systolic array, i.e., the core of TPU. Therefore, this paper implements systolic arrays as hardware accelerators for AxDNNs. Next, we characterize the fault resilience of AxDNNs as follows. \\

\noindent \textbf{(a) Accelerator specifications:} We vary the degree of approximation in systolic arrays with a fixed number of faulty MAC units and, in another case, vary the number of faulty MAC units inside the systolic array of TPU and compare the corresponding accuracy loss with faults inside the tiles of GPU. The faults are injected only in one tile at one execution cycle. This analysis helps us understand how different components of accelerators contribute to the fault resilience of AxDNNs. \\

\noindent \textbf{(b) Fault-bit representations:} For a detailed fault resilience analysis, we analyze how much a fault affects the approximate multipliers output bits leading to significant accuracy loss. In particular, we studied the impact of stuck-at 0 and stuck-at 1 faults at different output bit positions (LSB to MSB) of all approximate multipliers in AxDNNs by analyzing their classification accuracy. \\

\noindent \textbf{(c) Model specifications:} We further extend our analysis to layer-wise fault resilience of AxDNNs. Since a fault in each tile or a layer can propagate to all downstream layers \cite{condia2021combining} therefore, this is a special case of stuck-at faults used for this detailed analysis. Our analysis also includes investigating the impact of different activation functions and the width of layers on the fault resilience of AxDNNs. Lastly, we explore the fault energy trade-offs by analyzing the approximate systolic array's energy consumption and fault resiliency in each AxDNN. Such trade-off exploration results in multiple energy-aware and fault-resilient and non-resilient knobs.

The usability of our evaluation methodology is highly dependent on the ease of customizing AxDNNs and simulating the fault injection in the hardware accelerators. In this paper, customizing AxDNNs for fault injection is made simple by determining the mapping of their computations on the TPU and GPU. Note that our evaluation methodology can be used for any type of AxDNNs.

\section{Proposed fault-aware weight retuning for fault mitigation (Fal-reTune)}
\label{sec:mitigation}

In this section, we propose a novel Fal-reTune mitigation method for enhancing the fault resilience of AxDNNs. Fal-reTune initially employs pruning of the weights mapped to the faulty MAC units. The fault locations are determined through post-fabrication tests on an AxDNN chip. This initial step is similar to bypassing a MAC unit in AccDNNs using a multiplexer at the hardware level in systolic arrays \cite{zhang2018analyzing}. With the bypass path enabled, the contribution of the faulty MAC units to the column sum is skipped. However, bypassing a single faulty MAC unit may result in pruning multiple pre-trained weights due to the reuse of systolic arrays in the data processing. This problem is typically solved through retraining in the state-of-the-art fault mitigation techniques \cite{zhang2018analyzing, abdullah2020salvagednn, zhang2019fault} for AccDNN models. However, these techniques cannot be directly adopted for fault mitigation in AxDNNs as retraining AxDNNs increases the overall runtime of the training process. For example, training an AxDNN with 50 layers having arbitrary approximate components takes 89 days \cite{mrazek2016design}. This is due to the fact that the simulation of approximate arithmetic components does not usually scale well, and their training is thus slowed down by one order of magnitude for CPU and in three orders for GPU, \cite{mrazek2019alwann}. 

Let us recall that our AxDNNs are built by replacing the accurate multipliers with approximate multipliers in the inference phase. However, their training phase uses accurate multipliers. This gives us a hint on how to mitigate faults in AxDNNs, e.g., by retraining the accurate counterpart of AxDNNs, and then retuning the un-pruned weights (inspired by weight retuning in \cite{mrazek2019alwann}) in approximate inference. Based on this idea, our proposed Fal-reTune mitigation retunes the unpruned weights using the weight mapping function $f_M$ that is pre-calculated offline for each multiplier $M$ offline. For each weight $w$, $f_M$ determines a new weight $w$' that minimizes the sum of absolute differences between the output of the approximate and accurate multiplication overall activations $a$ $\subset$ $A$ ($A$ refers to activations matrix) as follows:

\begin{equation}
    f_M (w) = \arg\max_{w' \in W'} \sum\limits_{a \in A} | M (a, w') - a \cdot w |,
\end{equation}

where $W$' refers to the approximate weight matrix. For an accurate multiplier, $w'$ = $w$. Algorithm 1 delineates the steps involved in the proposed Fal-reTune mitigation method. Lines 1-2 prune the pre-trained weights of AccDNN mapped to the faulty PEs in systolic arrays. The retraining of the accurate counterpart of AxDNN starts from Line 3. Line 4 updates the un-pruned weights through back-propagation in retraining. Line 5 sets the weights mapped to faulty PEs as zero at the end of each retraining epoch. Line 7 performs the retuning of the un-pruned weights in approximate inference. Finally, Fal-reTune checks the classification accuracy of the approximate inference using the updated weights. The Fal-reTune algorithm returns the retraining accuracy to the user.

\SetInd{0.5em}{0.5em}
\begin{algorithm}[!t] \small
\caption{Fal-reTune Mitigation Algorithm}
\label{alg:robust}
\DontPrintSemicolon
\algsetup{linenosize=\small}

\SetKwInOut{Input}{Inputs}\SetKwInOut{Output}{Outputs}
\Input{(i) pre-trained weights: wts; (ii) training data: trData; (iii) test data: tsData; (iv) fault maps: fmaps; (v) max retraining epochs: trEpochs; (vii) approximate multiplier: mx;  (viii) accuracy threshold: accThresh;}
\Output{Accuracy: acc;}
\begin{algorithmic}[1]
\STATE ind = FindPrunedWeightsIndices (fmaps, wts) \\
//Find indices of pruning weights from fault maps 

\STATE pWts = SetPrunedWeightsToZero(ind, wts) \\
//Assign zero to the pruning weights at above indices

\FOR{epochs = 0 : trEpochs - 1}{
    \STATE nWts = UpdateWeights (pWts, trData) \\
    //Update weights with backpropagation
    \STATE nWts = SetUpdatedWeightsToZero(nWts, ind) \\
    //Assign zero to all pruning weights using indices in Step 1
}\ENDFOR

\STATE nWts' = ApproxWeightMapping(nWts, mx) \\
//Update weights using the weight mapping scheme

\STATE acc = CheckInferenceAccuracy(nWts', tsData) \\
//Check inference accuracy using new weights 

\RETURN acc;
\end{algorithmic}
\end{algorithm}

\section{Results and Discussions}
\label{sec:results}

\begin{figure*}[!b]
     \centering
     \begin{subfigure}[b]{0.245\textwidth}
         \centering
         \includegraphics[width=1\textwidth]{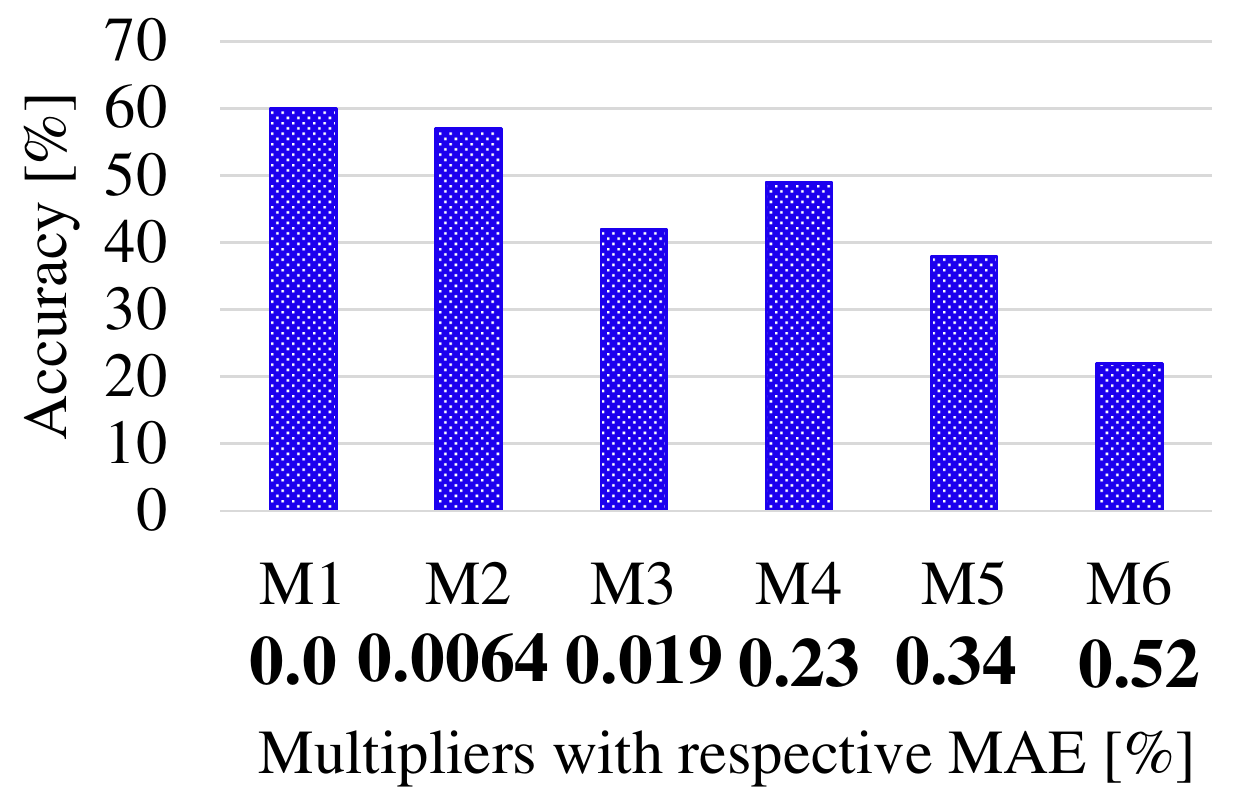}
         \caption{MP-tanh classification}
         \label{subfig:degree_a}
     \end{subfigure}
    \vspace{-0.09in}
    \hfill
     \begin{subfigure}[b]{0.245\textwidth}
         \centering
         \includegraphics[width=1\textwidth]{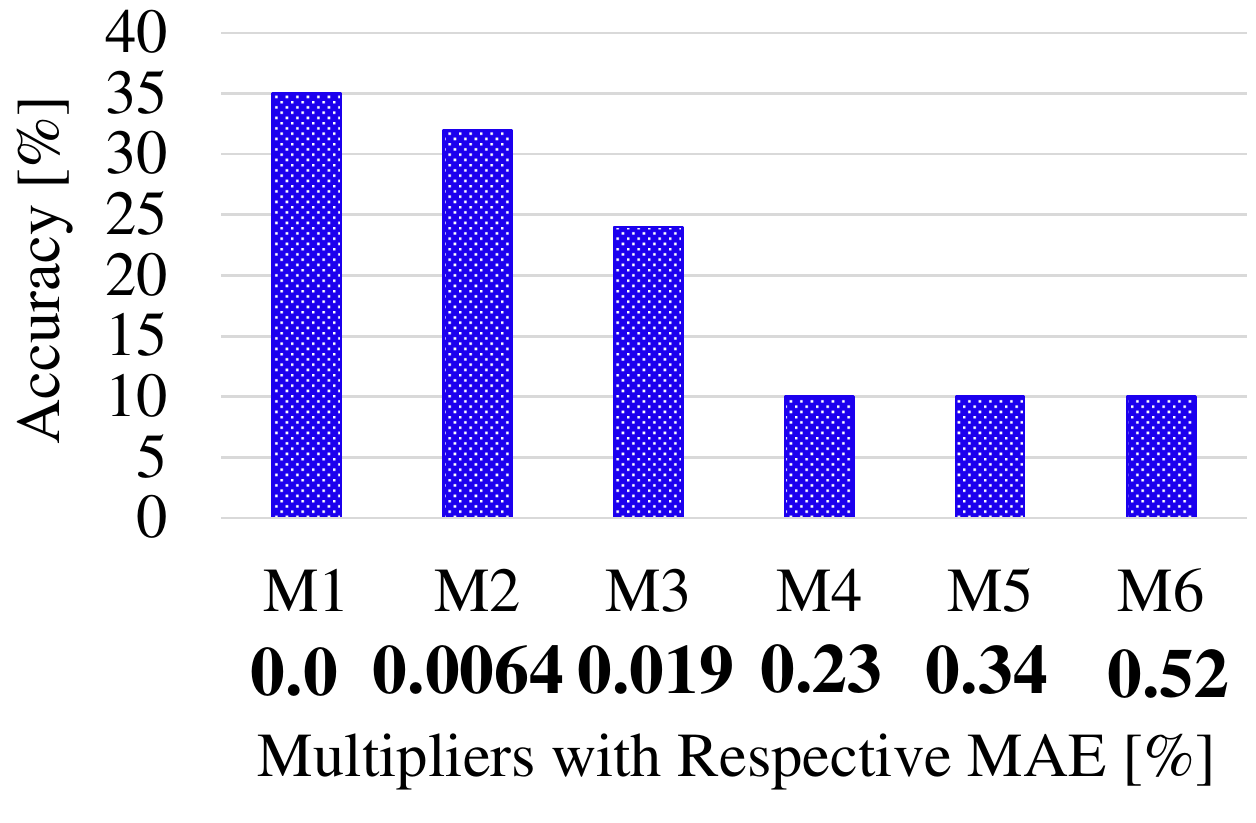}
         \caption{CC-Alx Classification}
         \label{subfig:degree_b}
     \end{subfigure}     
    \vspace{-0.09in}
     \hfill
     \begin{subfigure}[b]{0.245\textwidth}
         \centering
         \includegraphics[width=1\textwidth]{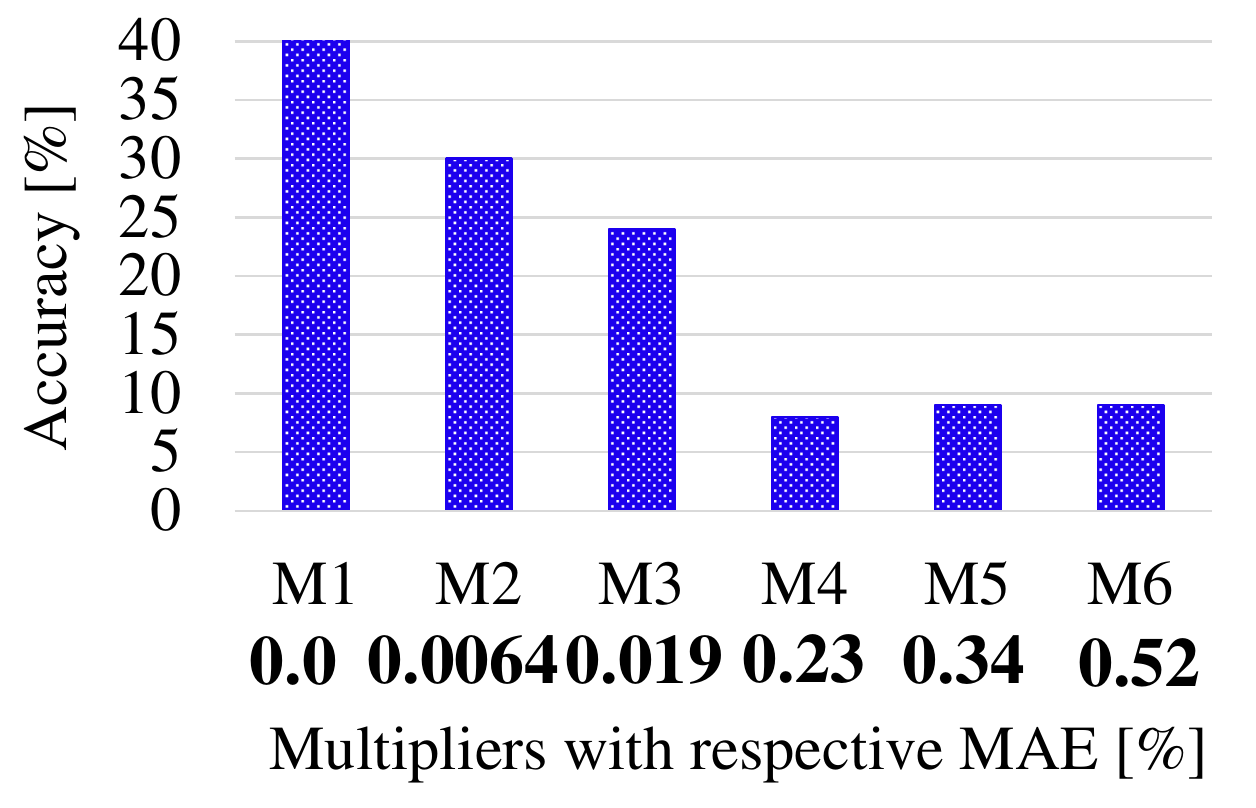}
         \caption{FP-tanh Classification}
         \label{subfig:degree_c}
     \end{subfigure}     
    \vspace{-0.09in}
    \hfill
     \begin{subfigure}[b]{0.245\textwidth}
         \centering
         \includegraphics[width=1\textwidth]{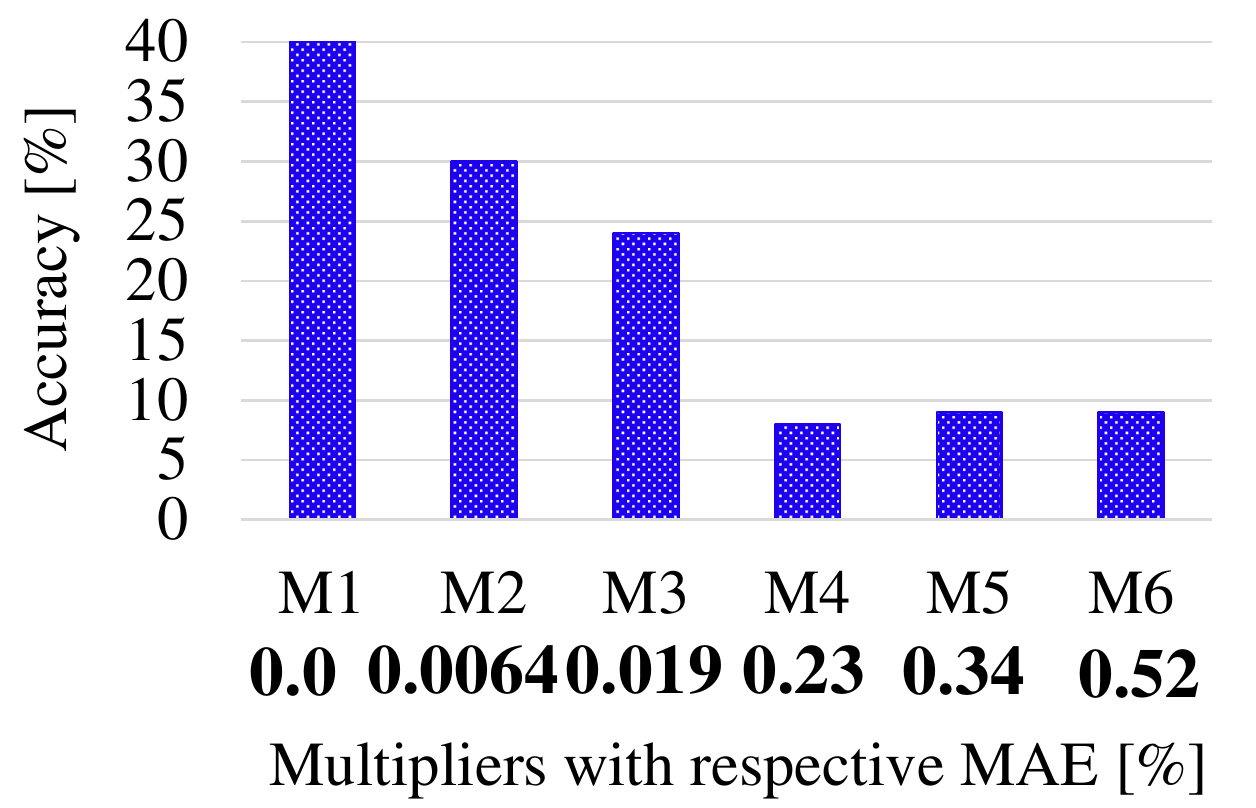}
         \caption{CC-VGG Classification}
         \label{subfig:degree_d}
     \end{subfigure} 
\vspace{0.2in}
\caption{Impact of stuck-at faults on approximate multipliers M\textit{n} based AxDNNs when they are injected in 16\% MAC units of TPU. The MAE of each multiplier in AxDNNs is written at the bottom.}
\vspace{-0.01in}
\label{fig:degree}
\end{figure*}

This section discusses the experimental setup and results for extensive fault resilience analysis and fault mitigation in AccDNN and AxDNNs using the Evoapprox8b \cite{mrazek2017evoapproxsb} library. In this section and onwards, the accurate multiplier is denoted as M1, and approximate multipliers refer to KVB, KX2, L2D, KTY, and L1G (denoted as M2-M11) signed approximate multipliers obtained from the Evoapprox8b library. Note that the approximation error is quantified using the error metric mean average error (MAE) of these multipliers.

\subsection{Datasets}
We use three image classification datasets i.e., MNIST \cite{cohen2017emnist}, Fashion-MNIST \cite{xiao2017fashion} and CIFAR10 \cite{krizhevsky2009learning}. The DNN research community widely uses these datasets for evaluating the performance of DNNs~\cite{marchisio2021feeca, marchisio2020descnet, putra2020fspinn} in embedded platforms. MNIST is a grayscale image dataset that recognizes handwritten digits (ranging from zero to nine). On the other hand, Fashion-MNIST consists of grayscale article images with labels from 10 different classes, such as boots, shirts, etc. These datasets contain 60,000 training and 10,000 validation images with 28x28 pixels. CIFAR10 is a colored image dataset used for object recognition. It includes 50,000 training and 10,000 validation colored images, with labels from different classes, of 32x32 pixels.

\begin{table}[!t]
\caption{{Classification Accuracy of Benchmarked Accurate Multi layer Perceptron (MLP) and Convolutional Neural Networks (CNN)}}
\label{tab:accuracy}
\centering
\begin{tabular}{|l|l|l|l|}
\hline & \\[-2ex]
\cellcolor{gray!25}Datasets                       & \cellcolor{gray!25}Architecture  & \cellcolor{gray!25}Acronym    & \cellcolor{gray!25}Acc.\\ \hline & \\[-2ex]
\multirow{4}{*}{MNIST}         & MLP-tanh    & MP-tanh    & 96\%     \\ \cline{2-4}& \\[-2ex]
                               & MLP-softmax & MP-softmax & 96\%     \\ \cline{2-4} & \\[-2ex]
                               & CNN-Lenet5      & ML-base    & 97\%     \\ \cline{2-4} & \\[-2ex]
                               & CNN-Lenet5 wide & ML-wide    & 97\%     \\ \hline & \\[-2ex]
\multirow{2}{*}{\begin{tabular}[c]{@{}l@{}}Fashion-\\ MNIST\end{tabular}} & MLP-tanh    & FP-tanh    & 82\%     \\ \cline{2-4} & \\[-2ex]
                               & MLP-softmax & FP-softmax & 81\%     \\ \hline & \\[-2ex]
\multirow{2}{*}{CIFAR10}       & CNN-Alexnet     & CC-Alx    & 82\%     \\ \cline{2-4} & \\[-2ex]
                               & CNN-VGG-11      & CC-VGG     & 82\%     \\ \hline
\end{tabular}
\vspace{-0.1in}
\end{table}

\subsection{Model Configurations}

We perform our fault resilience analysis on three different basic DNN models. For MNIST, we define a baseline MLP architecture with layer configuration 784-256-256-256-10 and generate two variants, i.e., MP-tanh and MP-softmax. The MP-tanh and MP-softmax differ in tanh and softmax activation functions only. We also use Lenet-5 as a base model ML-base and generate its variant ML-wide by widening the layer size to 5x5x20, 5x5x40, 100, and 10 for Conv1, Conv2 and FC1, and FC2 layers. Here, Conv1 and Conv2 denote the first two convolutional layers, and FC1 and FC2 denote two fully connected layers in LeNet-5. Likewise, for Fashion-MNIST, we generate a baseline MLP architecture, base (FP), with layer configuration 784-512-512-512-10 and generate two MLP variants: FP-tanh and FP-softmax having different activation functions. We also examine an off-the-shelf network CNN model, AlexNet, denoted as FC-Alex. Finally, for CIFAR10, we employ both AlexNet and VGG-11, i.e., CC-Alex and CC-VGG, respectively. The classification accuracy of accurate MP-tanh, MP-softmax, FP-tanh, FP-softmax, CC-Alex, CC-VGG, ML-base, and ML-wide is provided in Table \ref{tab:accuracy}. Their approximate counterparts are generated by simply replacing the accurate multipliers with approximate multipliers in the inference phase. In the following sections, we discuss the application of these models in this paper. 

\begin{figure*}[!t]
     \centering
     \vspace{-0.01in}
     \begin{subfigure}[t]{0.245\textwidth}
         \centering
         \includegraphics[width=1\textwidth]{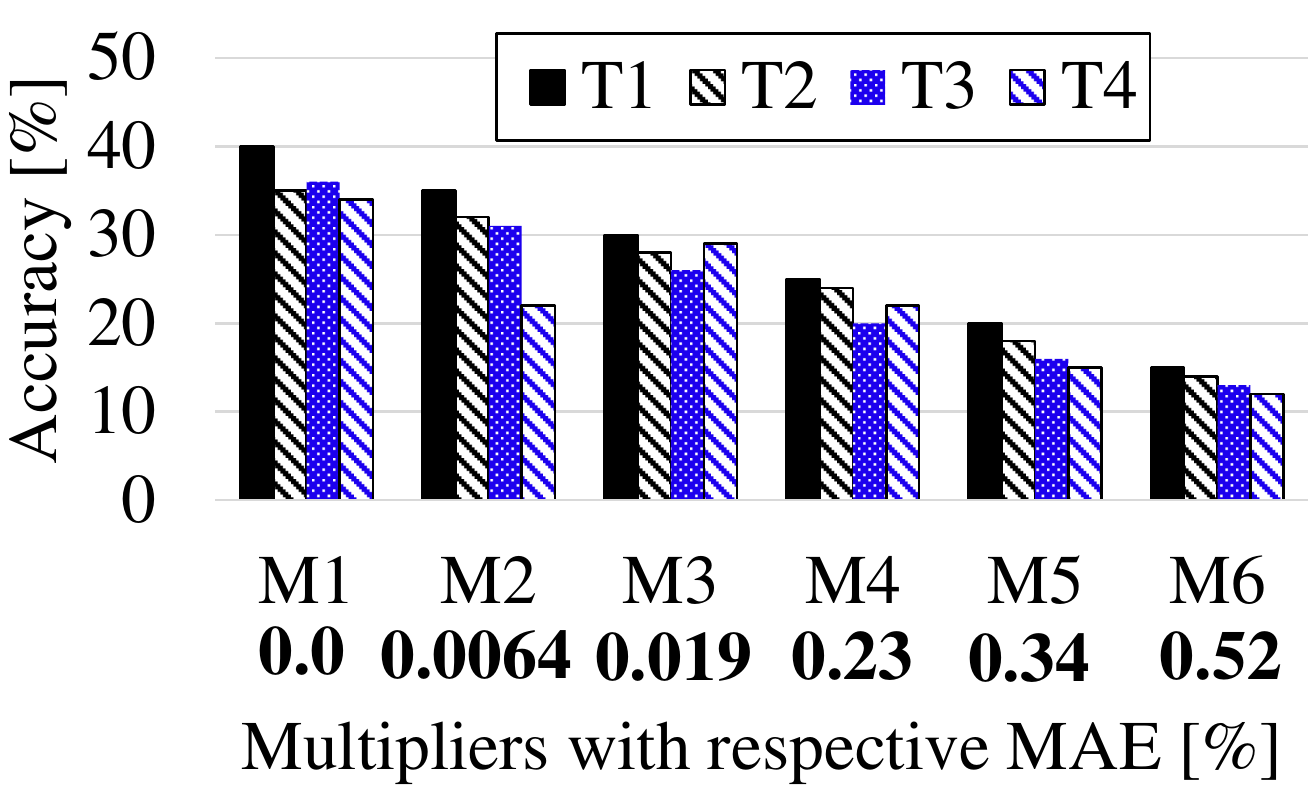}
         \caption{MP-tanh classification}
         \label{subfig:accelerator_a}
     \end{subfigure}
    \vspace{-0.09in}
    \hfill
     \begin{subfigure}[t]{0.245\textwidth}
         \centering
         \includegraphics[width=1\textwidth]{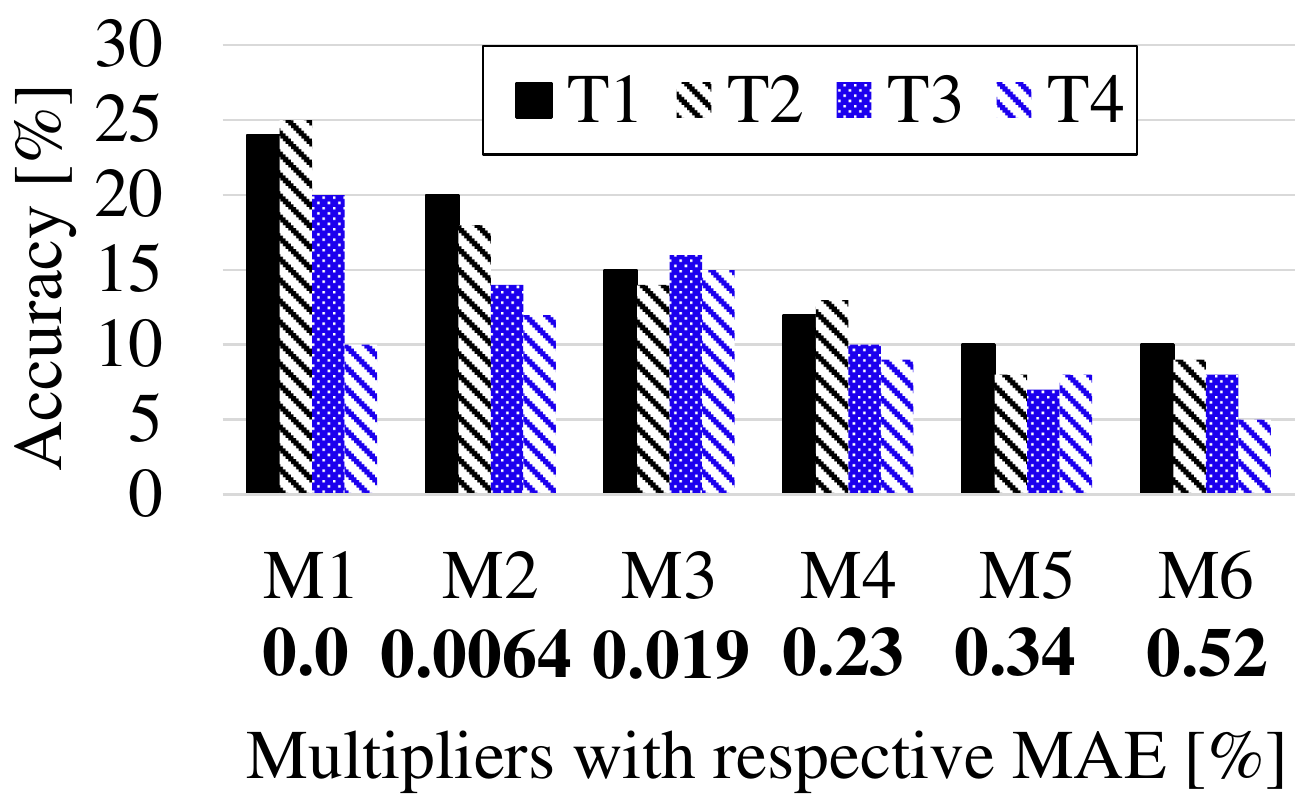}
         \caption{CC-Alx Classification}
         \label{subfig:accelerator_b}
     \end{subfigure}     
    \vspace{-0.09in}
     \hfill
     \begin{subfigure}[t]{0.245\textwidth}
         \centering
         \includegraphics[width=1\textwidth]{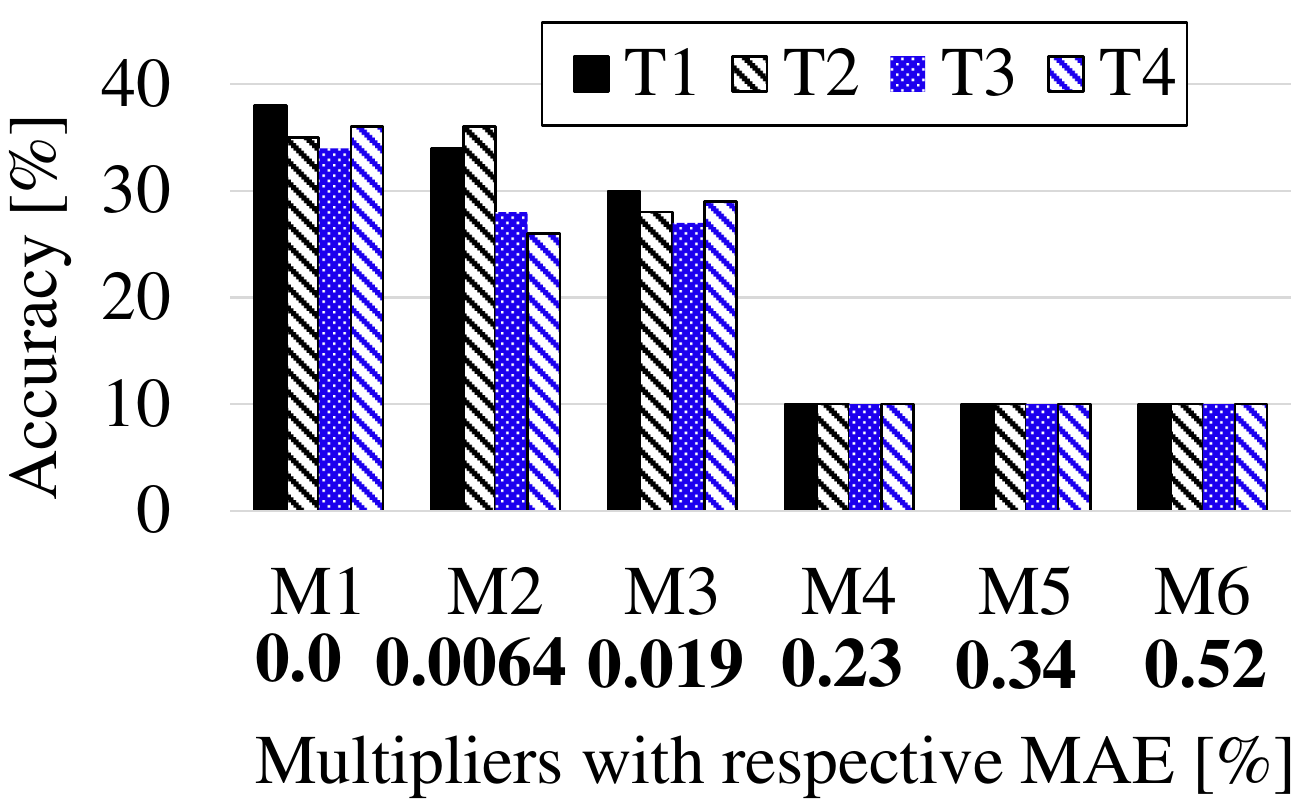}
         \caption{FP-tanh Classification}
         \label{subfig:accelerator_c}
     \end{subfigure}     
    \vspace{-0.09in}
    \hfill
     \begin{subfigure}[t]{0.245\textwidth}
         \centering
         \includegraphics[width=1\textwidth]{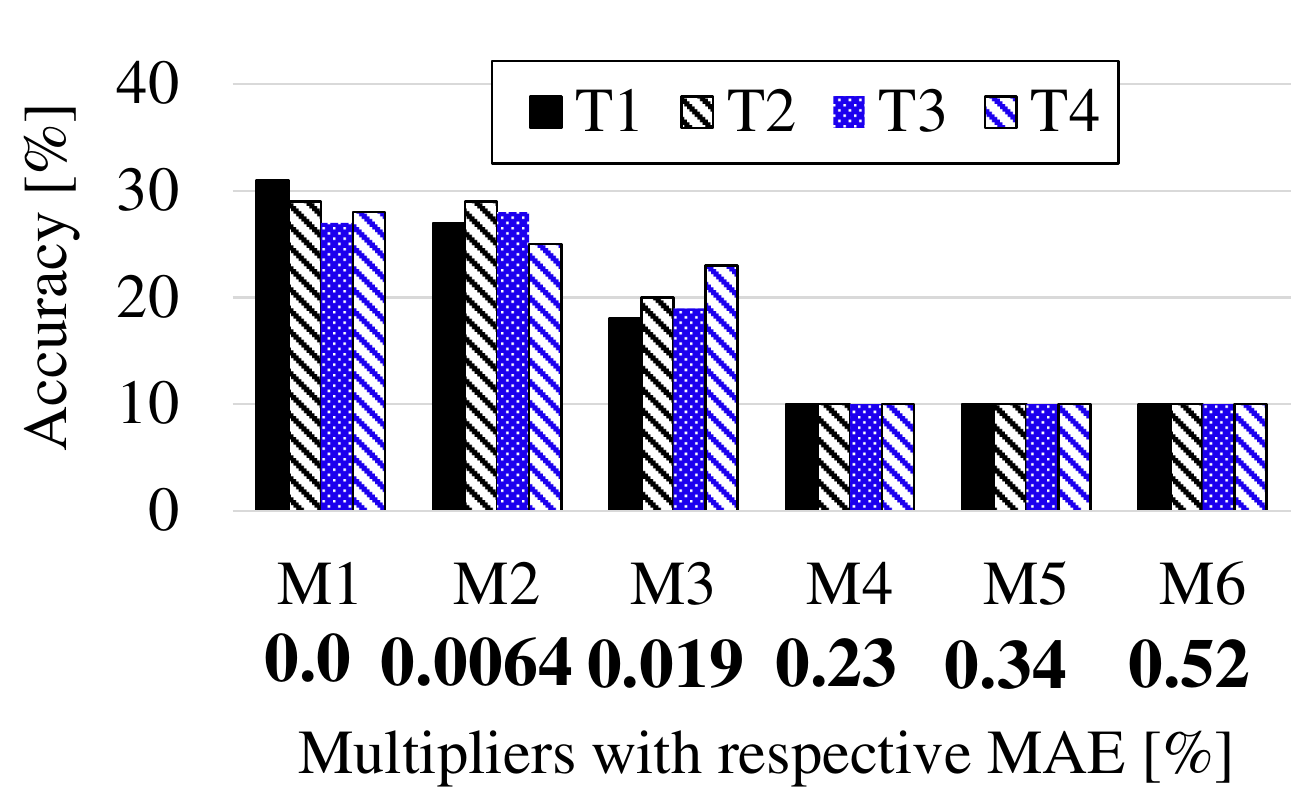}
         \caption{CC-VGG Classification}
         \label{subfig:accelerator_d}
     \end{subfigure} 
\vspace{0.25in}
\caption{Impact of stuck-at faults on approximate multipliers M\textit{n} based AxDNNs when they are injected in different tiles of GPU. The MAE of each multiplier in AxDNNs is written at the bottom.}
\label{fig:accelerator}
\end{figure*}

\begin{figure*}[!t]
     \centering
     \vspace{-0.01in}
     \begin{subfigure}[t]{0.245\textwidth}
         \centering
         \includegraphics[width=1\textwidth]{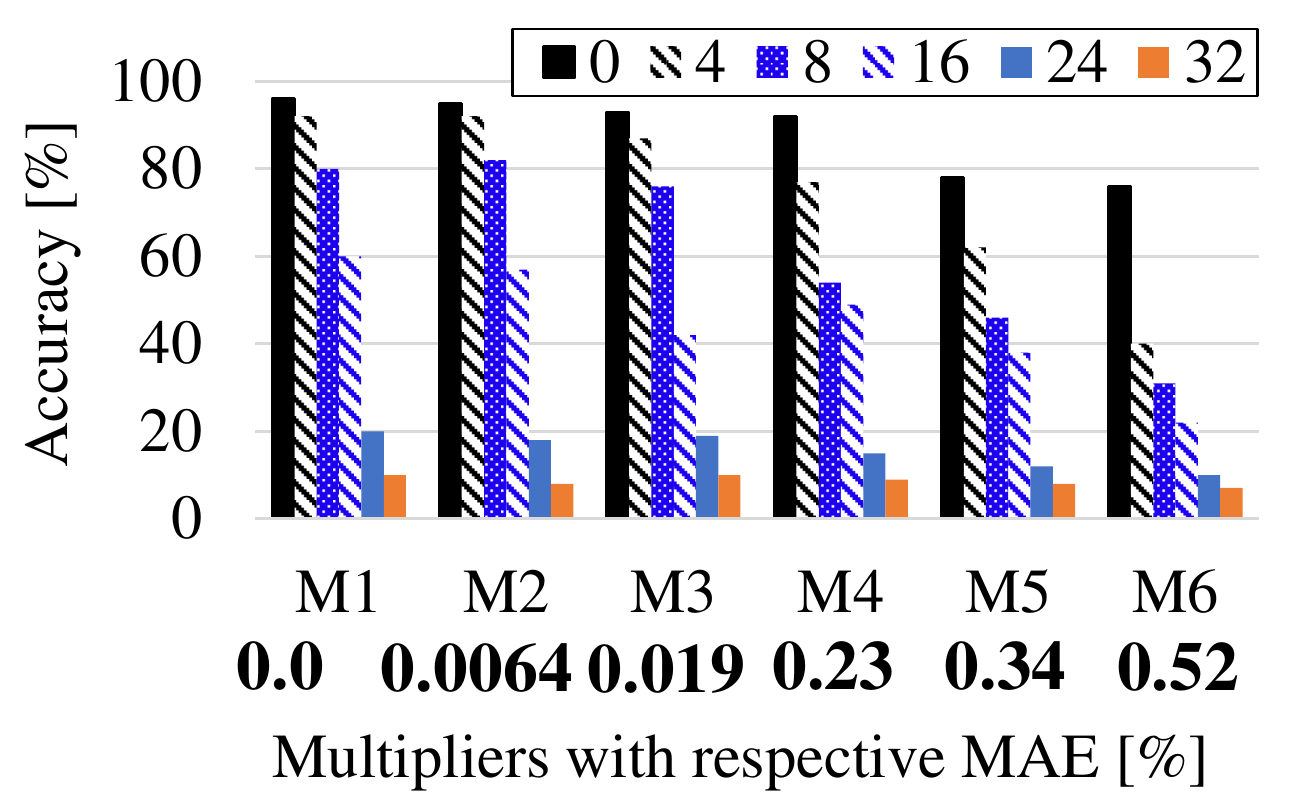}
         \caption{MP-tanh classification}
         \label{subfig:accelerator_2a}
     \end{subfigure}
    \vspace{-0.09in}
    \hfill
     \begin{subfigure}[t]{0.245\textwidth}
         \centering
         \includegraphics[width=1\textwidth]{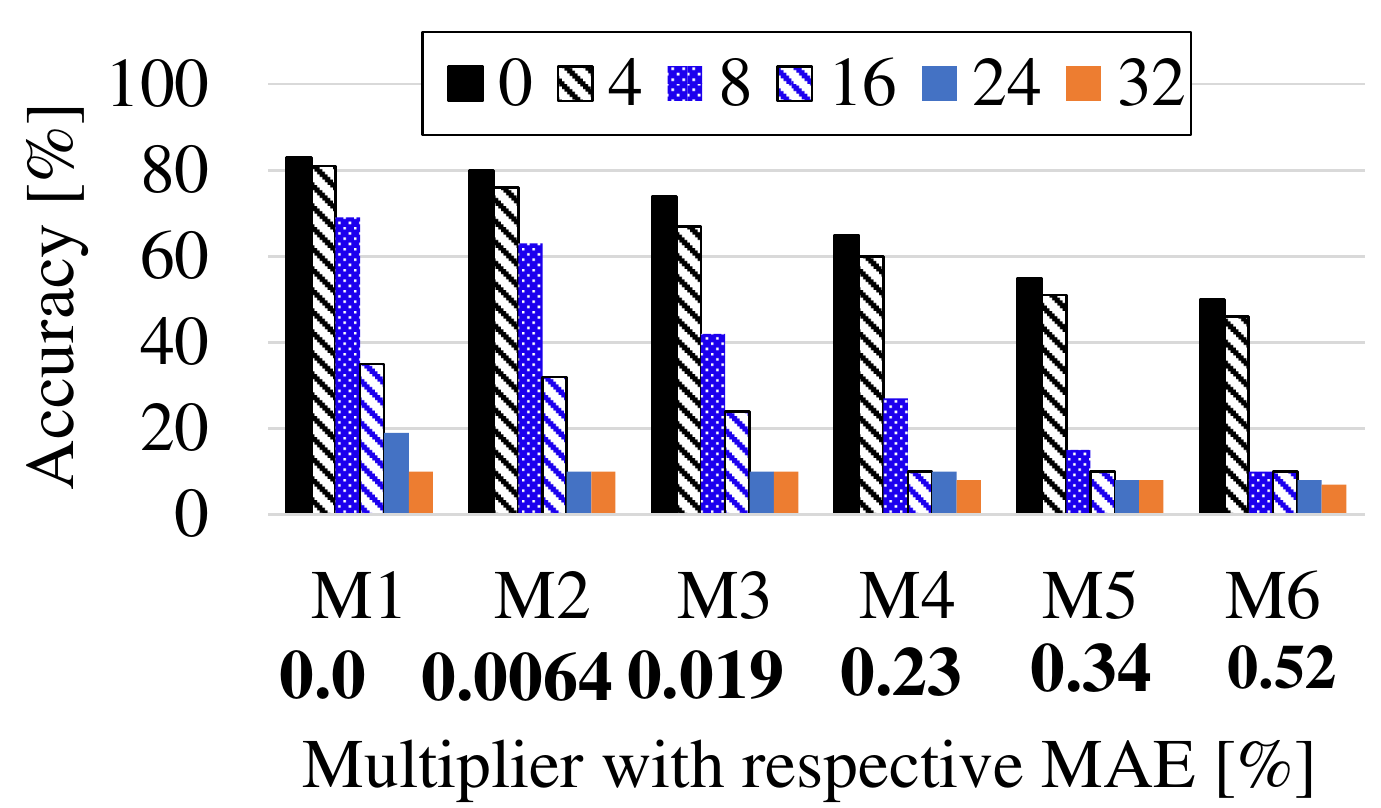}
         \caption{CC-Alx Classification}
         \label{subfig:accelerator_2b}
     \end{subfigure}     
    \vspace{-0.09in}
     \hfill
     \begin{subfigure}[t]{0.245\textwidth}
         \centering
         \includegraphics[width=1\textwidth]{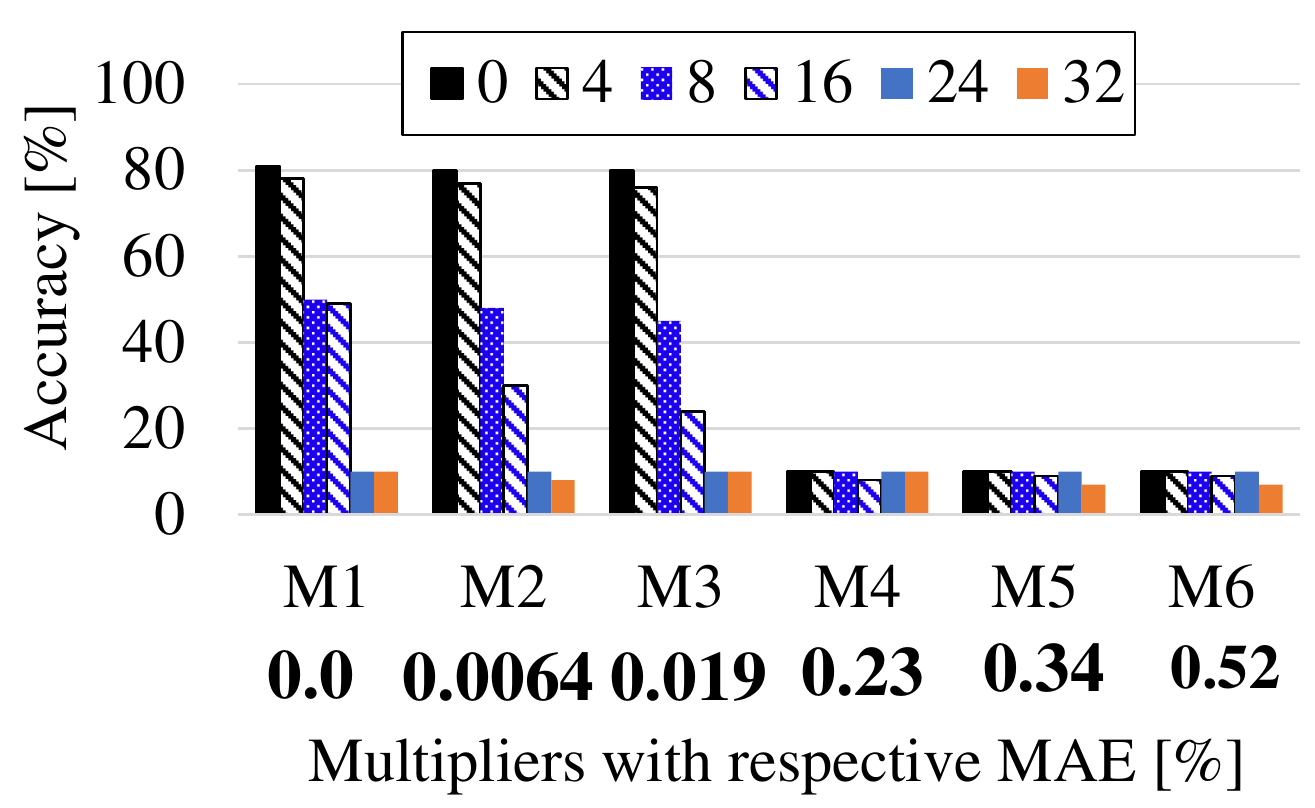}
         \caption{FP-tanh Classification}
         \label{subfig:accelerator_2c}
     \end{subfigure}     
    \vspace{-0.09in}
    \hfill
     \begin{subfigure}[t]{0.245\textwidth}
         \centering
         \includegraphics[width=1\textwidth]{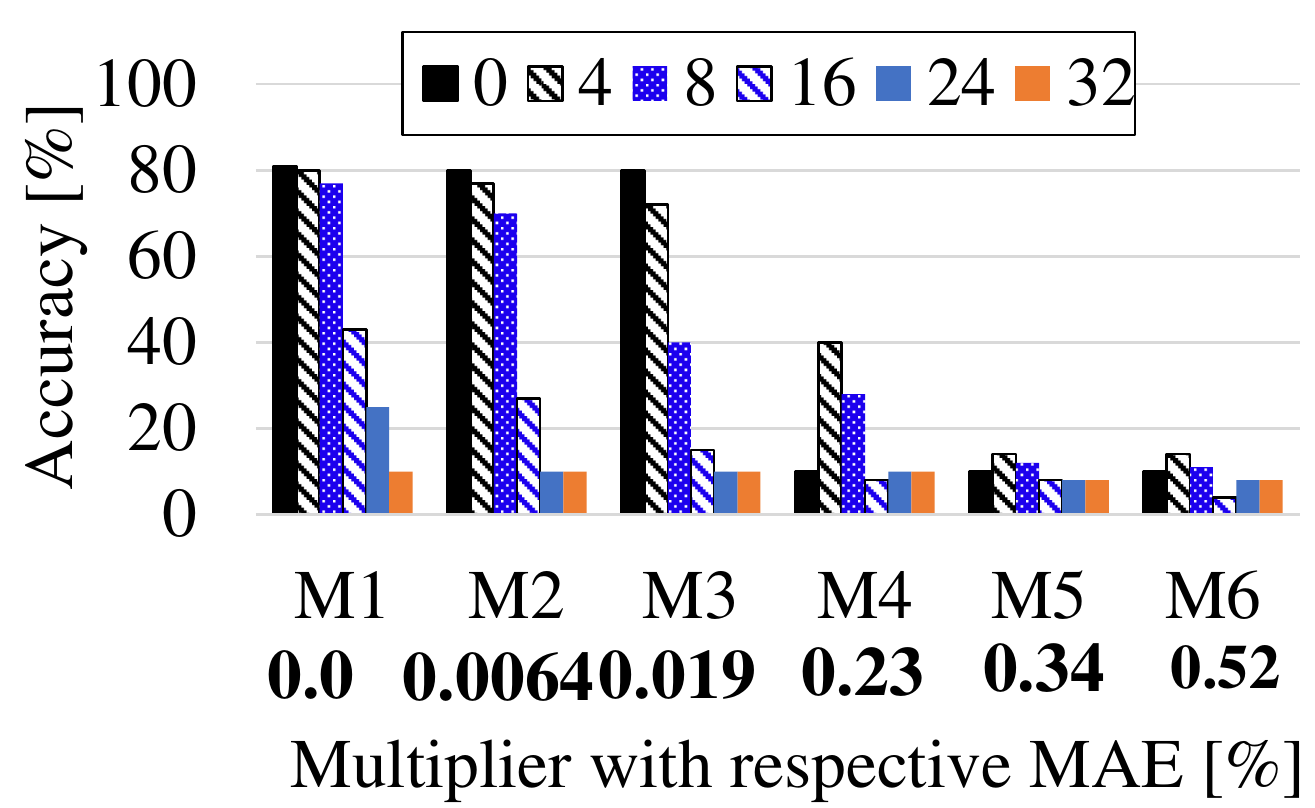}
         \caption{CC-VGG Classification}
         \label{subfig:accelerator_2d}
     \end{subfigure} 
\vspace{0.25in}
\caption{Impact of stuck-at faults on approximate multipliers M\textit{n} based AxDNNs when they are injected in 0\% , 4\% , 8\% , 16\% , 24\%  and 32\% MAC units of TPU. The MAE of each multiplier in AxDNNs is written at the bottom.}
\label{fig:accelerator2}
\end{figure*}

\subsection{Fault Injection and Energy Estimation}
Our fault resilience analysis framework systematically injects stuck-at 0 and stuck-at 1 faults in AccDNN and AxDNN models and quantifies their impact using the accuracy loss metric, i.e., drop in classification accuracy with fault injection in a model as compared to the fault-free counterpart. It is focused only on the stuck-at faults in the data-path and ignores faults in the memory components as they can be mitigated by using error correction codes easily. Also, it ignores faults in the control logic that consume an insignificant fraction of the design. In both AccDNNs and AxDNNs, the faults can be either inside or outside a computational unit of the systolic array, i.e., inside a multiplier or in the output of a multiplier. However, a fault inside a multiplier may or may not get masked by the other parts of the logic in the multiplier. We, therefore, mimic the situation where the faults are not masked and propagated to the output of the multiplier. This is achieved by injecting the faults in the output of the multiplier. Note the output of an approximate multiplier is different (approximated output) from an accurate multiplier. Hence, a stuck-at fault in the output of an approximate multiplier will have a different impact when compared to the same fault in the output of an accurate multiplier. In this paper, we implement the fault injection analysis using Python 3.73 and Tensorflow 2.2, which supports CUDA 10.1 for accelerating computations by using GPUs. Our experiments run on the NVIDIA GeForce RTX 2080 Ti GPU with Intel Core i9-10900kF operating at 3.06 GHz with 32 GB RAM. 

For analyzing the energy consumption of approximate multipliers, we first developed the behavioral design of each approximate multiplier and simulate them in Xilinx to generate a Value Change Dump (VCD) file. We used this file and behavioral design as inputs to the Synopsys Design Compiler for latency and power estimation, with a OSU FreePDK 45nm standard cell library, after logic synthesis. The latency and power reports are then used for determining the energy of the design.

\subsection{Fault Analysis with Accelerator specifications}
\label{subsesc:acceleratorspec}
In this section, we investigate how approximation and type of underlying accelerator characterize fault resilience in AxDNNs.

\subsubsection{Impact of the degree of approximation}
\label{subsubsec:approximationdegree}
To examine the impact of different degrees of approximation on the fault resilience of the underlying hardware, we injected stuck-at faults in 16\% MAC units of TPU by perturbing the MSB of their approximate multiplier's output in a 256x256 systolic array. In particular, we generated a fault map and then mapped it to the MAC units in the systolic array of approximate TPU running AxDNN. A fault map refers to a matrix that keeps track of the faulty MAC units in a systolic array. The size of the fault map is the same as the size of the systolic array and the elements of the fault map have a one-to-one mapping to the multiplier-and-accumulator (MAC) units of the systolic array. Note, we selected the MSB for fault injection in this experiment to analyze the worst case analysis as the fault in MSB leads to significant accuracy loss \cite{mrazek2017evoapprox8b}. In Fig. \ref{fig:degree}, our results show that fault resilience decreases significantly when the approximation error is high in AxDNNs. For example, the fault injection in the M6 multiplier leads to only 22\% classification accuracy in MP-tanh, but the same fault injection in the M1 multiplier leads to 76\% classification accuracy (see Fig. \ref{subfig:degree_a}). This is due to the high MAE of the M6 multiplier (i.e., 0.52\%) compared to the M1 multiplier. The M1 multiplier is an accurate multiplier having zero MAE. The higher the approximation error is, the higher the MAE. We also observe that the fault injection in the M6 multiplier leads to only 10\% classification accuracy in CC-Alex; this accuracy is also very low when compared to the same fault injection in the M1 multiplier that leads to 35\% classification accuracy (see Fig. \ref{subfig:degree_b}). Also, M4, M5, and M6 approximate multipliers undergo significant accuracy loss in the presence of faults; MP-tanh only undergoes comparatively less accuracy loss with these multipliers. The reason is that MP-tanh has comparatively high prediction accuracy without faults, as listed in Table \ref{tab:accuracy}. Similar trends are observed with FP-tanh (see Fig. \ref{subfig:degree_c}) and CC-VGG (see Fig. \ref{subfig:degree_d}) classification.

\subsubsection{Impact of the accelerator type}
\label{subsubsec:acceleratortype}
We also characterize how a fault in the core of TPU and GPU affects the fault resilience of AxDNNs. For this purpose, we assume that the stuck-at faults are located in the core tiles of GPU and only 40\% of one tile is
damaged at one time. We chose the number 40\% as we did not observe any considerable impact of faults when less than 40\% of a tile was damaged during our experiments. Fig. \ref{fig:accelerator} shows that such fault injection significantly affects both AccDNNs and AxDNNs. Our results demonstrate that the classification accuracy stays in the range 40\% to 12\% and 5\% to 25\% in MP-tanh (see Fig. \ref{subfig:accelerator_a}) and CC-Alx (see Fig. \ref{subfig:accelerator_b}) running on a GPU with faults in the tiles; though, the approximate multiplier with higher MAE are affected more by the faults as compared to the ones with lower MAE. A similar trend is observed with FP-tanh (see Fig. \ref{subfig:accelerator_c}) and CC-VGG (see Fig. \ref{subfig:accelerator_d}) classification. This is due to the same reason as discussed in Section \ref{subsubsec:approximationdegree}.

The GPU tiles are often mapped to a systolic array for more robust parallel matrix multiplications. However, a fault in the systolic array can propagate to all layers of AxDNNs. Let us consider the faults injected in the different percentage numbers of MAC units of approximate TPU running AxDNNs. The results in Fig. \ref{fig:accelerator2} show that even a small number of faulty MAC units can lead to significant accuracy degradation in AxDNNs. For example, 8\% faulty MAC units can drop the classification accuracy of MP-tanh and CC-Alx from 90\% to 52\% and 63\% to 25\%, respectively, when they are built using the M4 approximate multiplier (having 0.23\% MAE). However, when MP-tanh and CC-Alx are built using M1 multiplier, these faulty MAC units lead to 96\% to 52\% (see Fig. \ref{subfig:accelerator_2a}) and 82\% to 69\% (see Fig. \ref{subfig:accelerator_2b}) accuracy drop only. Similar trends are observed with FP-tanh (see Fig. \ref{subfig:accelerator_2c}) and CC-VGG (see Fig. \ref{subfig:accelerator_2d}) classification. Here, the accuracy drop is severe in TPU because it is repeatedly used for layer execution, and therefore, all layers suffer from the fault injection in a PE. Furthermore, an increase in the percentage number of faulty MAC units leads to an even more significant accuracy drop e.g., 32\% faulty MAC units lead to 8\% classification accuracy only in both AccDNNs and AxDNNs.

\subsubsection{Impact of the accelerator size}
\begin{figure}[!h]
     \centering
     \vspace{-0.05in}
     \begin{subfigure}[t]{0.24\textwidth}
         \centering
         \includegraphics[width=1\textwidth]{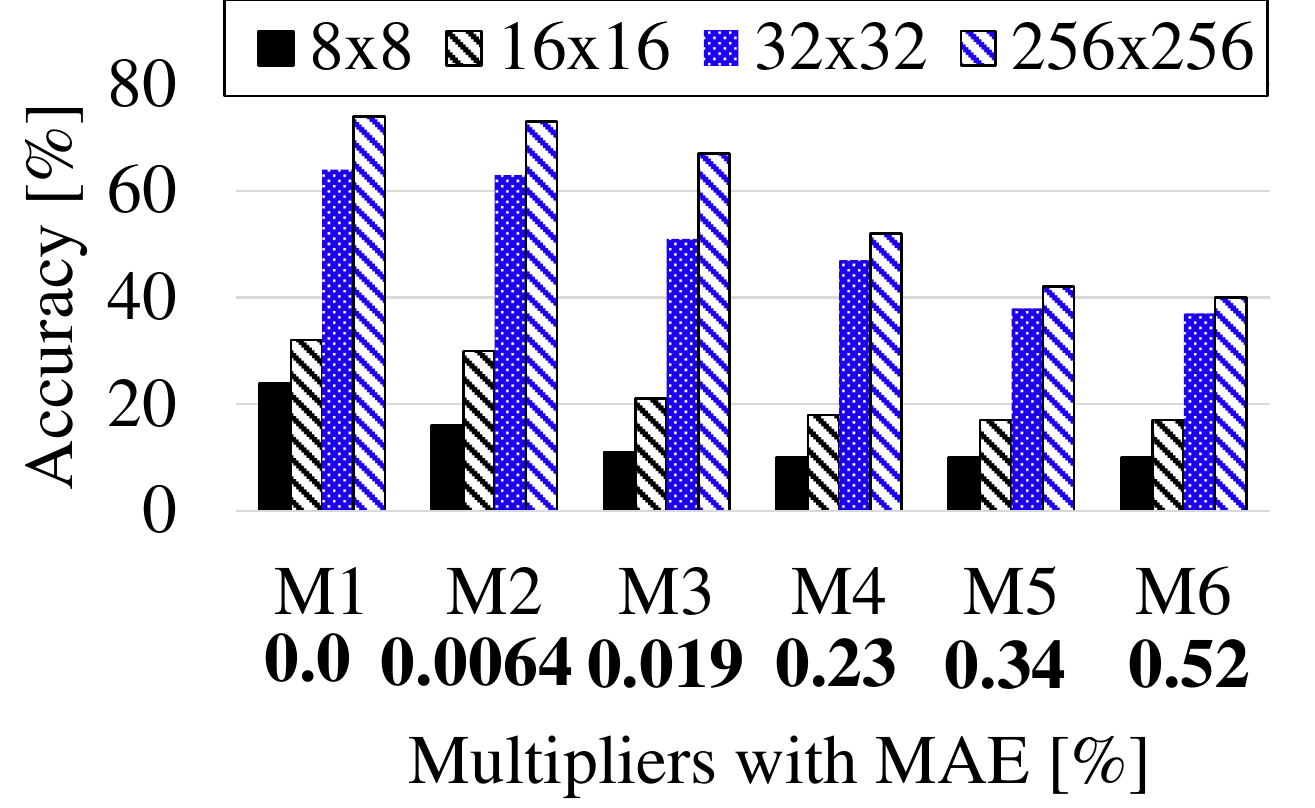}
         \caption{MP-tanh classification}
         \label{subfig:size_a}
     \end{subfigure}
    \vspace{-0.09in}
    \hfill
     \begin{subfigure}[t]{0.24\textwidth}
         \centering
         \includegraphics[width=1\textwidth]{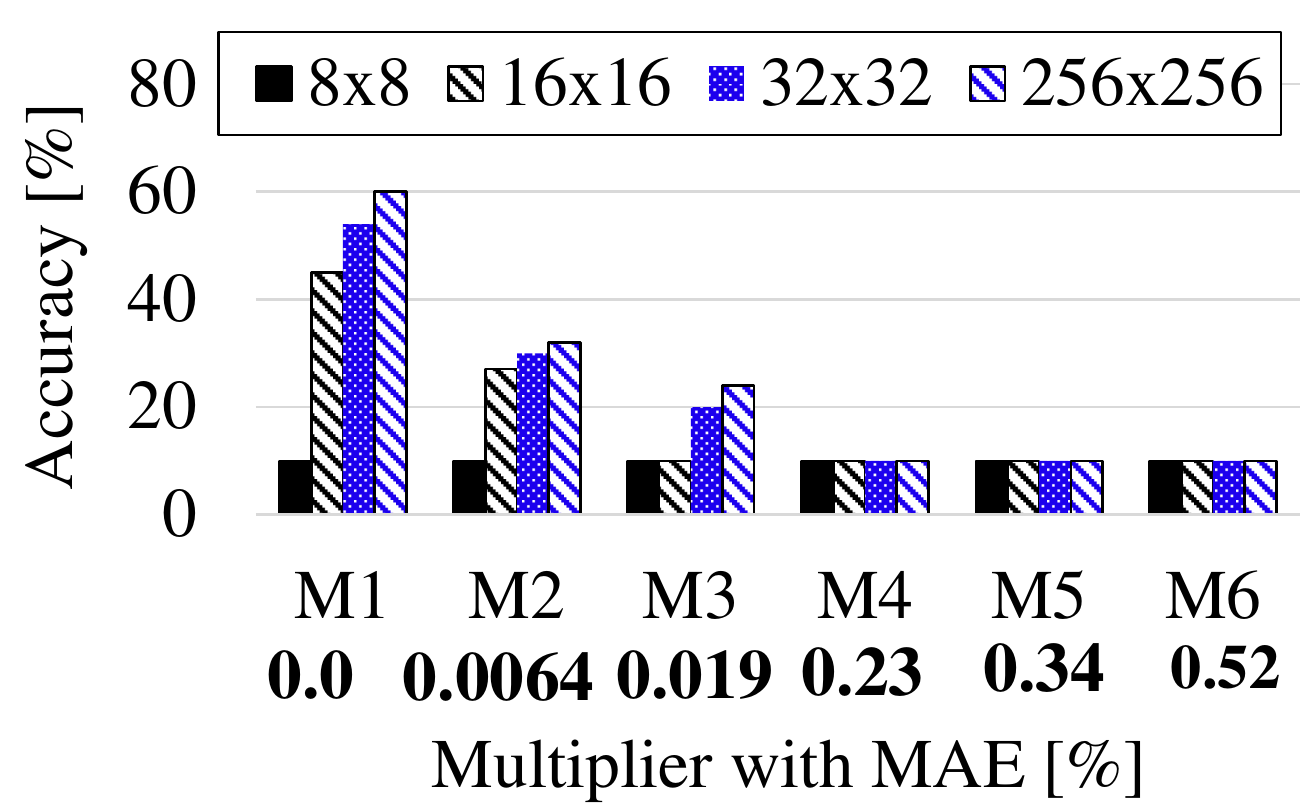}
         \caption{CC-Alx Classification}
         \label{subfig:size_b}
     \end{subfigure}   
\caption{Impact of stuck-at faults on AxDNNs (based on multipliers M1-M6) when they are injected in 16\% MAC units of systolic arrays having different sizes. The MAE of each multiplier in AxDNNs is written at the bottom.}
\label{fig:size}
\vspace{-0.1in}
\end{figure}

We further extend the above analysis by characterizing the fault resilience of AxDNNs with different sizes of the approximate systolic arrays. We vary the number of faulty PEs in them and run MP-tanh and CC-Alex. Our analysis in Fig. \ref{fig:size} shows that the smaller the size of the systolic array, the lower the fault resilience and vice versa. For example, an 8x8 systolic array has lower classification accuracy than a 256x256 systolic array in both MP-tanh and CC-Alx classification. This is because even though both smaller and larger AxDNN reuses the systolic array for processing the data in different layers, a smaller systolic array undergoes more execution cycles (due to insufficient mapping of the layer to the systolic array in one attempt).

\begin{figure*}[!t]
     \centering
     \vspace{-0.05in}
     \begin{subfigure}[t]{0.245\textwidth}
         \centering
         \includegraphics[width=1\textwidth]{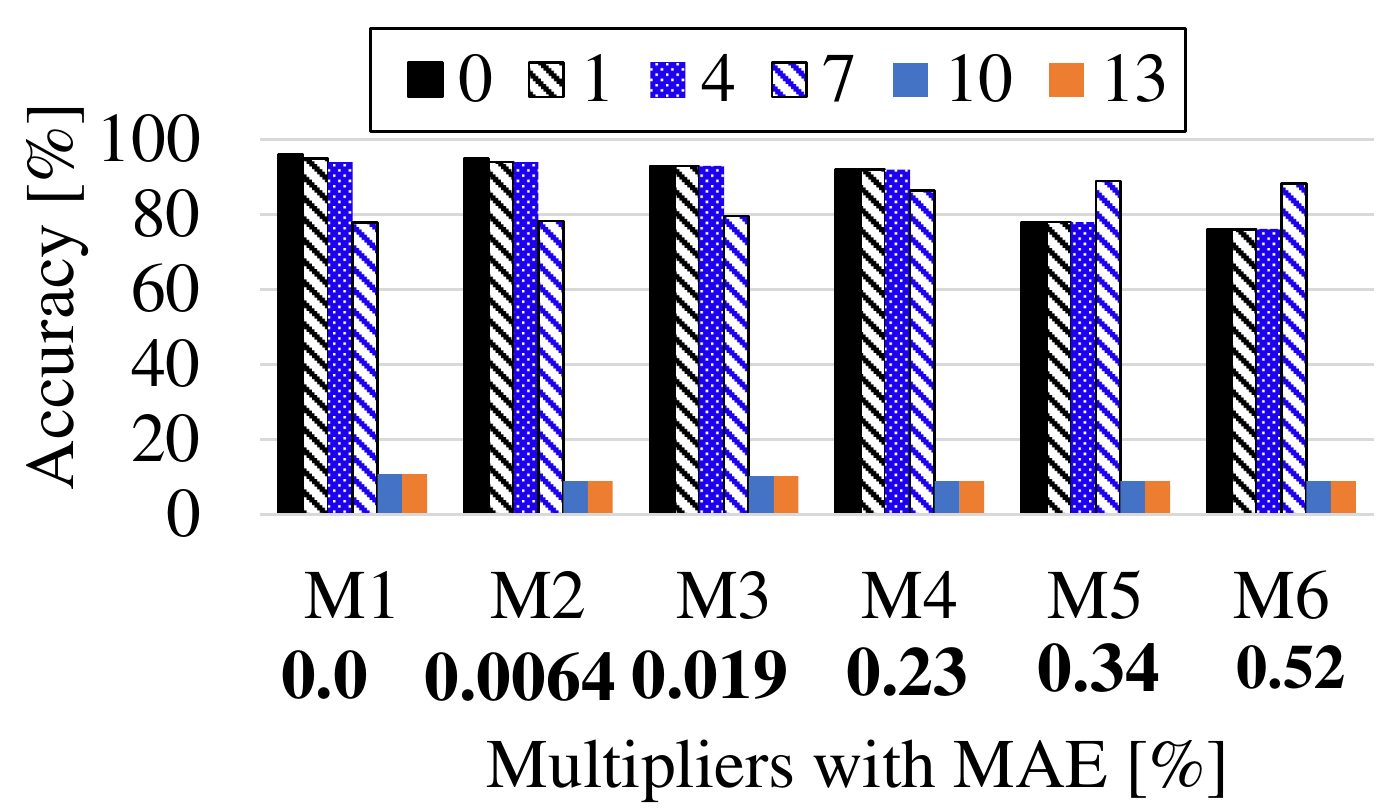}
         \caption{MP-tanh classification}
         \label{subfig:position_a}
     \end{subfigure}
    \vspace{-0.09in}
    \hfill
     \begin{subfigure}[t]{0.245\textwidth}
         \centering
         \includegraphics[width=1\textwidth]{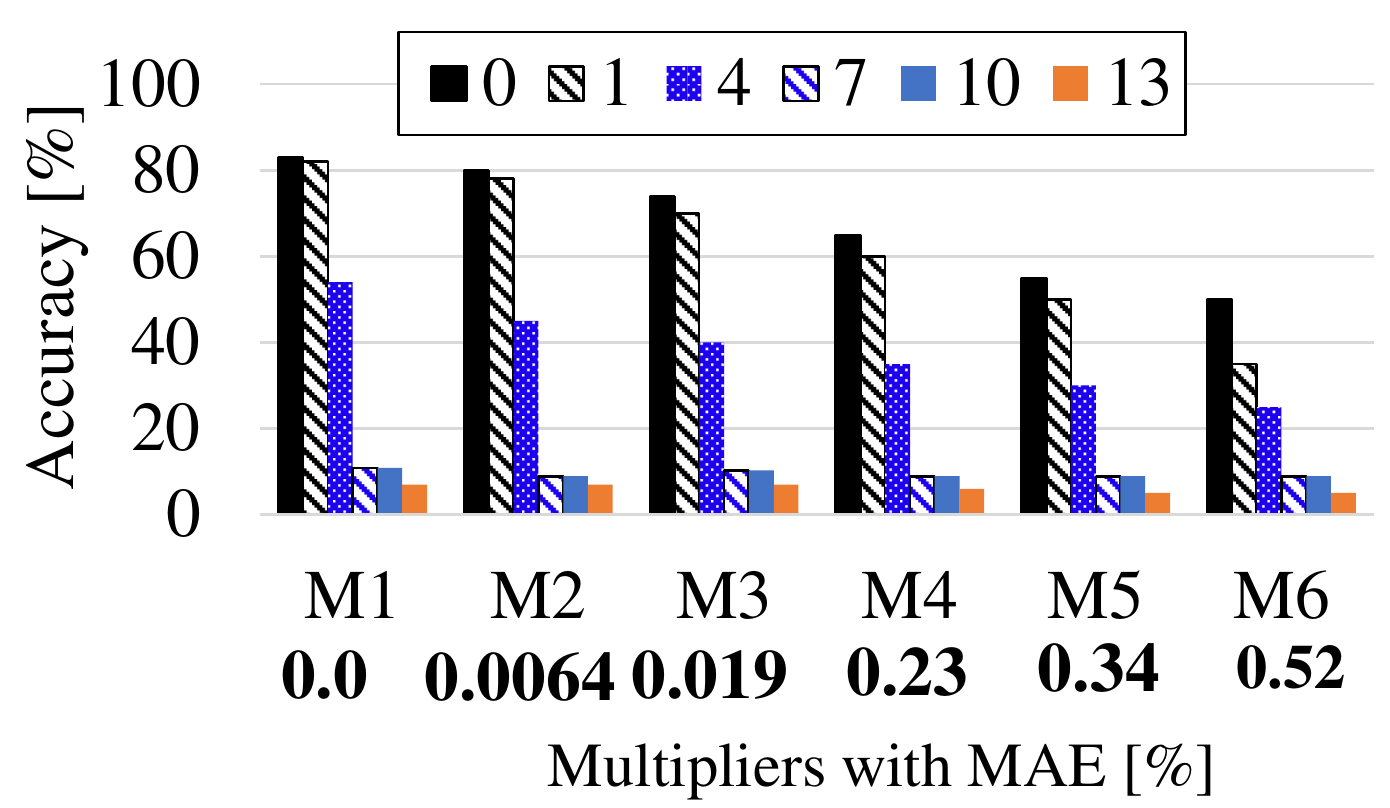}
         \caption{CC-Alx Classification}
         \label{subfig:position_b}
     \end{subfigure}     
    \vspace{-0.09in}
     \hfill
     \begin{subfigure}[t]{0.245\textwidth}
         \centering
         \includegraphics[width=1\textwidth]{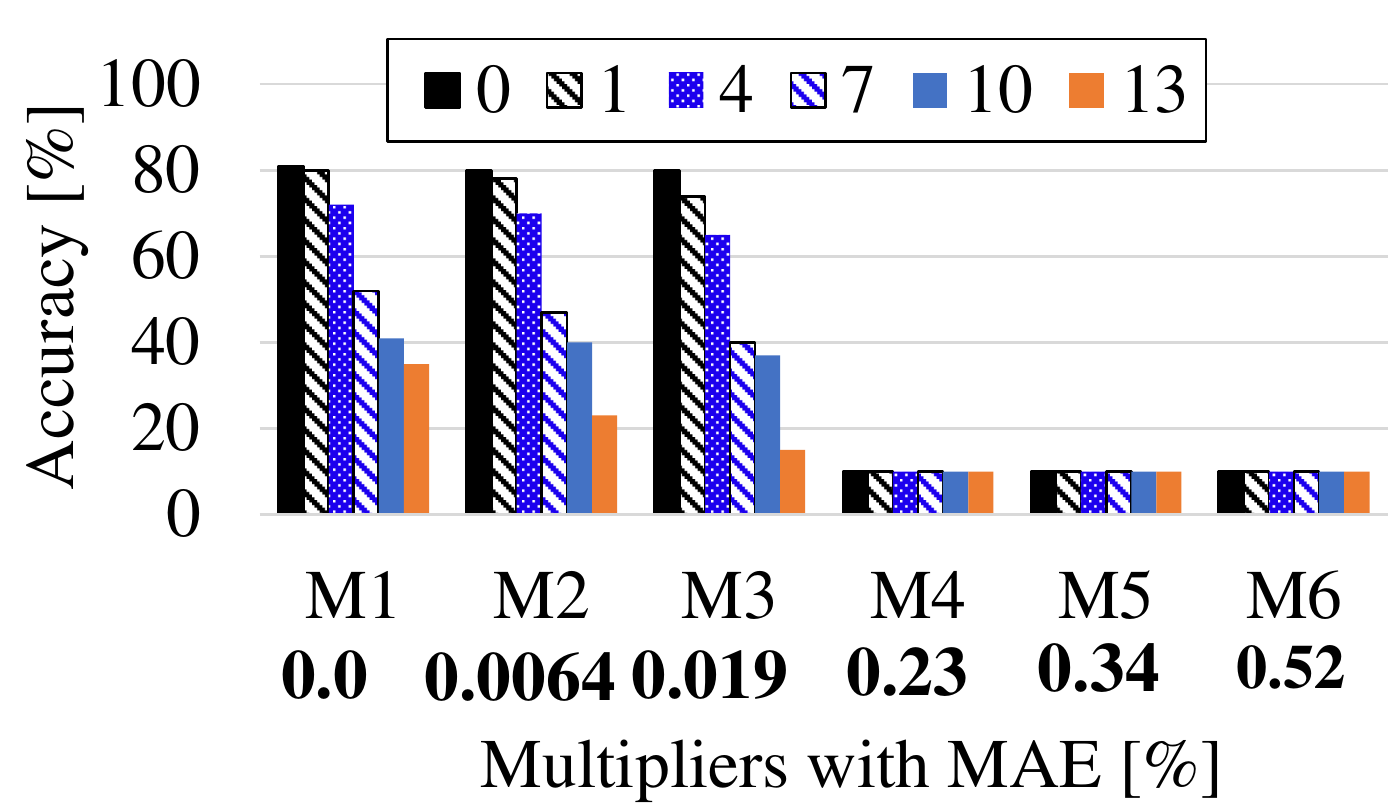}
         \caption{FP-tanh Classification}
         \label{subfig:position_c}
     \end{subfigure}     
    \vspace{-0.09in}
    \hfill
     \begin{subfigure}[t]{0.245\textwidth}
         \centering
         \includegraphics[width=1\textwidth]{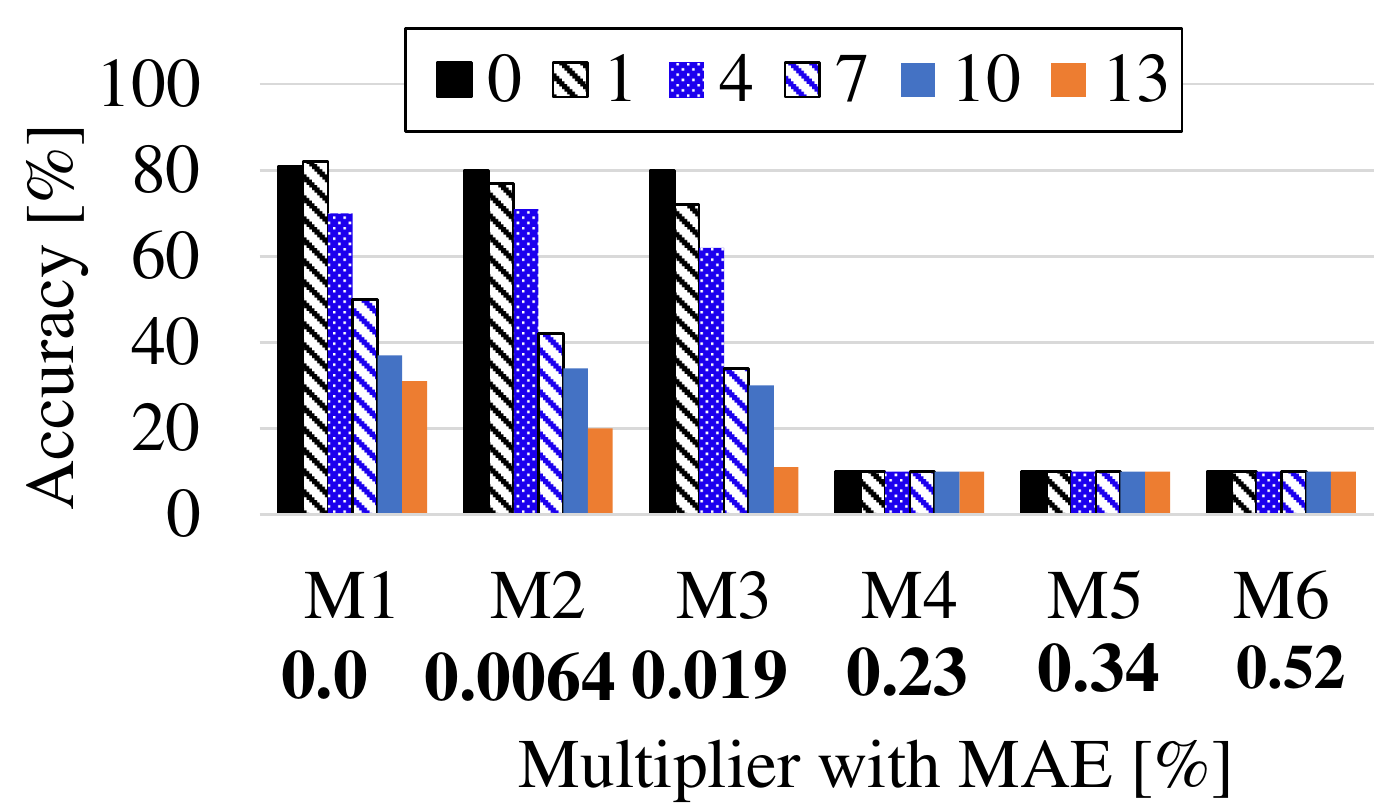}
         \caption{CC-VGG Classification}
         \label{subfig:position_d}
         0
     \end{subfigure} 
\vspace{0.25in}
\caption{Impact of stuck-at 1 faults on approximate multipliers M\textit{n} based AxDNNs when they are injected in 0, 1, 4, 7, 10 and 13 bit positions. The MAE of each multiplier in AxDNNs is written at the bottom.}
\label{fig:position}
\end{figure*}    

\begin{figure*}[!t]
     \centering
     \vspace{-0.05in}
     \begin{subfigure}[t]{0.24\textwidth}
         \centering
         \includegraphics[width=1\textwidth]{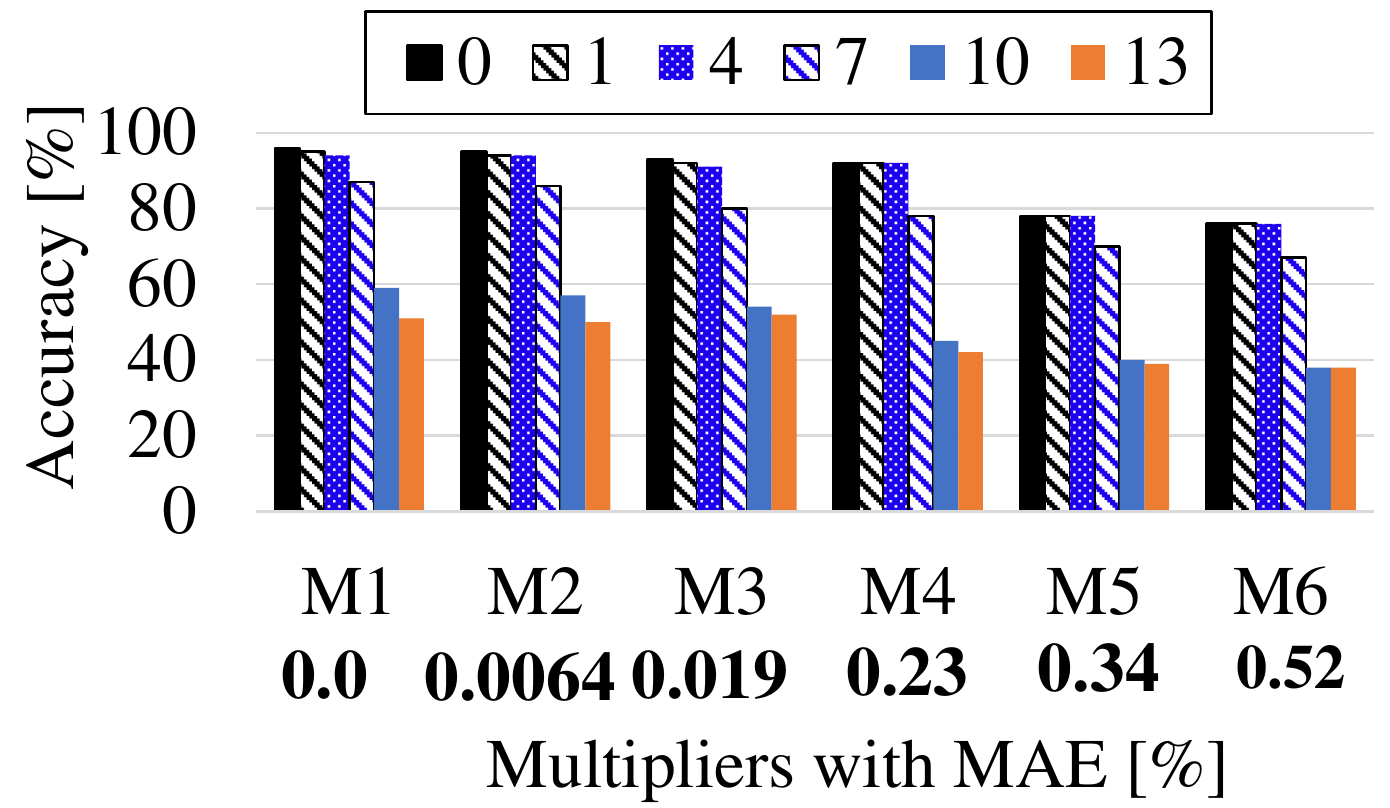}
         \caption{MP-tanh classification}
         \label{subfig:position_2a}
     \end{subfigure}
     \vspace{-0.09in}
     \hfill
     \begin{subfigure}[t]{0.24\textwidth}
         \centering
         \includegraphics[width=1\textwidth]{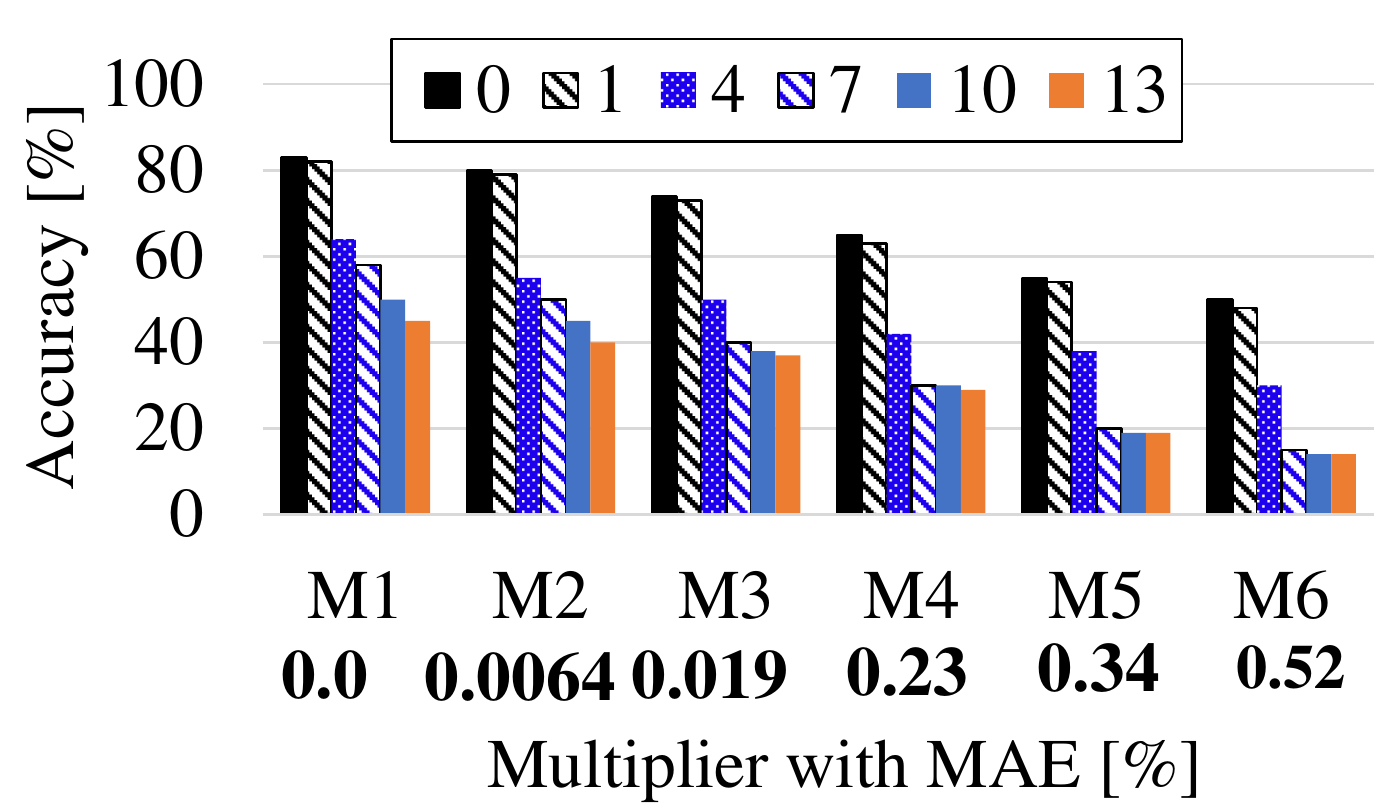}
         \caption{CC-Alx Classification}
         \label{subfig:position_2b}
     \end{subfigure}     
    \vspace{-0.09in}
    \hfill
     \begin{subfigure}[t]{0.24\textwidth}
         \centering
         \includegraphics[width=1\textwidth]{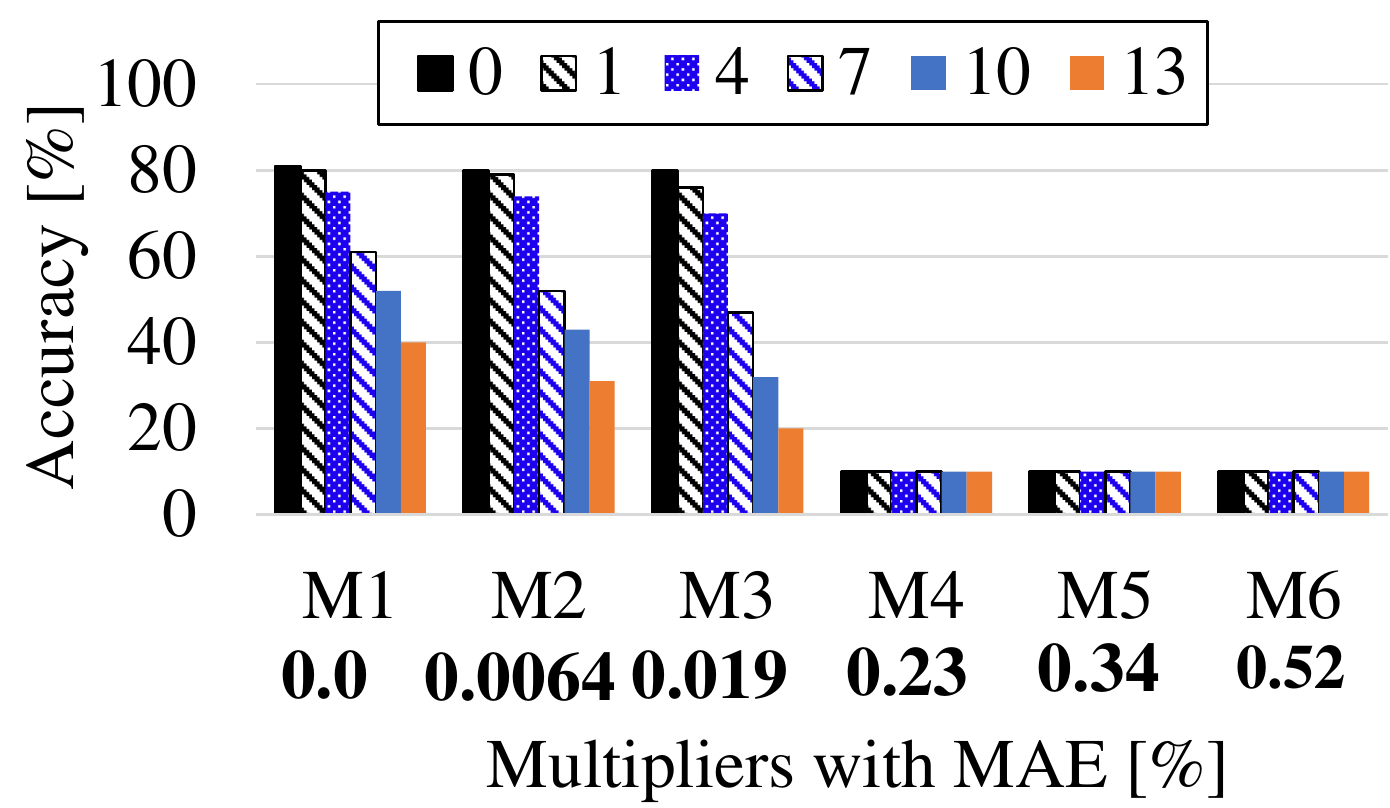}
         \caption{FP-tanh Classification}
         \label{subfig:position_2c}
     \end{subfigure}     
    \vspace{-0.09in}
    \hfill
     \begin{subfigure}[t]{0.24\textwidth}
         \centering
         \includegraphics[width=1\textwidth]{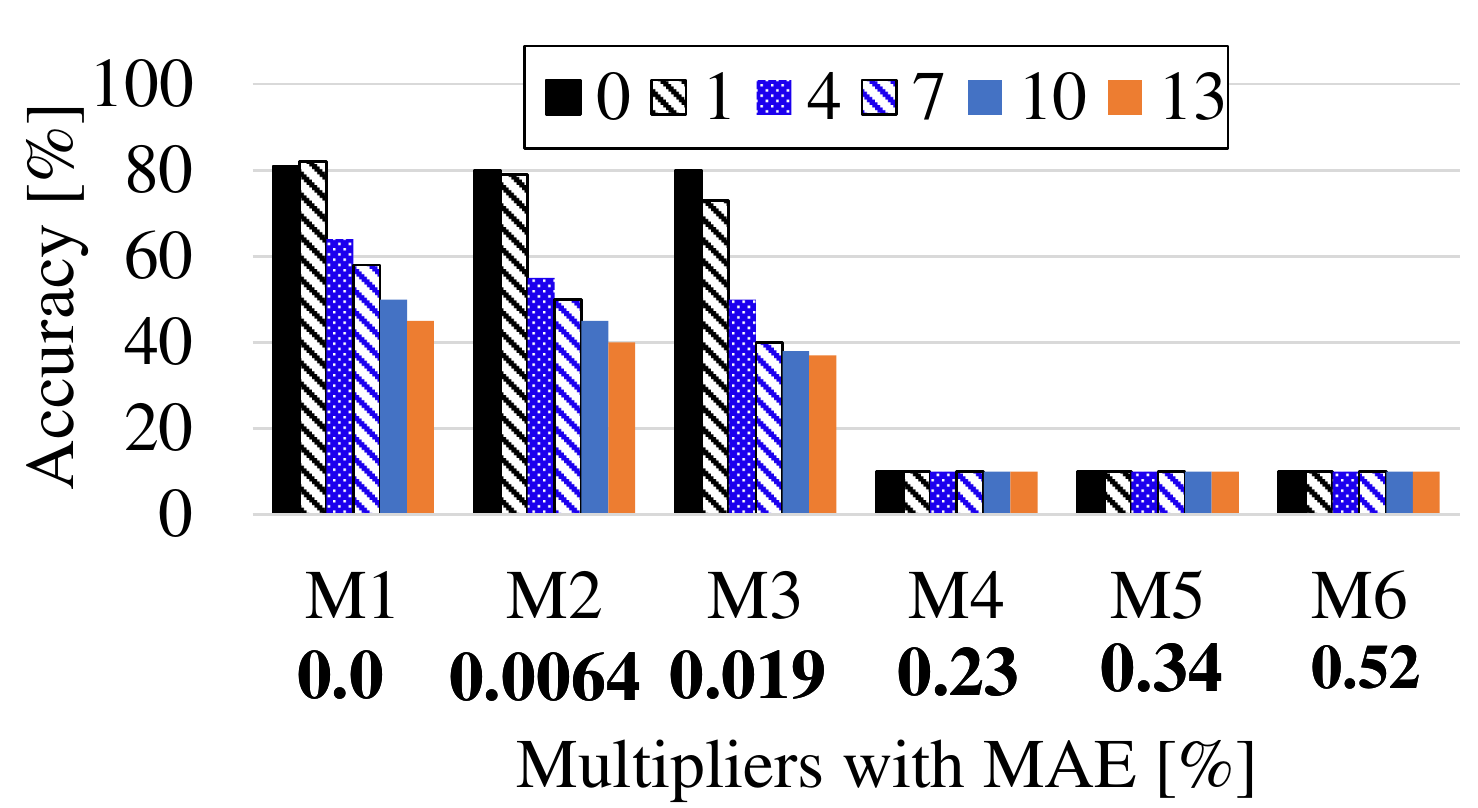}
         \caption{CC-VGG Classification}
         \label{subfig:position_2d}
     \end{subfigure}     
\vspace{0.3in}    
\caption{Impact of stuck-at 0 faults on approximate multipliers M\textit{n} based AxDNNs when they are injected in 0, 1, 4, 7, 10 and 13 bit positions. The MAE of each multiplier in AxDNNs is written at the bottom.}
\label{fig:position2}
\vspace{-0.15in}    
\end{figure*}

\subsection{Fault Analysis with fault-Bit Representations}
\label{subsec:bitwise}
In this section, we investigate that how faults in approximate bit representations characterize the fault vulnerability of AxDNNs.

\subsubsection{Impact of the fault-bit position}
We examine how much change in the approximation value is incurred by the fault bits that lead to significant accuracy loss in AxDNNs.

It is indeed obvious from the state-of-the-art works that the fault resilience of AccDNNs decreases from LSB to MSB. We investigate the similar trend in AxDNNs to demonstrate that they are relatively less fault resilient as compared to AccDNNs and this behavior is more clearly observable when we move towards lower order bits (LSBs). As shown in Fig. \ref{fig:position}, approximately 98\% to 100\% accuracy loss is observed with stuck-at 1 faults in MSB of accurate and approximate MP-tanh. On the other hand, the same fault in the LSB leads to an accuracy loss of approximately 0\% to 9\% in the approximate MP-tanh, which is comparatively higher than 1\% accuracy loss in accurate MP-tanh (see Fig. \ref{subfig:position_a}). Indeed, the fault resilience decreases from LSB to MSB in AxDNNs, similar to AccDNNs. For example, approximately 95\% to 100\% accuracy loss is observed with stuck-at 1 faults in MSB of accurate and approximate CC-Alx. On the other hand, the same fault in the LSB led to an accuracy loss of approximately 0\% to 15\% in the approximate CC-Alx, which is comparatively higher than 1\% accuracy loss in accurate CC-Alx (see Fig. \ref{subfig:position_b}). Similar trends are observed with FP-tanh (see Fig. \ref{subfig:position_c}) and CC-VGG (see Fig. \ref{subfig:position_d}) classification. Also, stuck-at 0 faults have a similar trend (see Fig. \ref{fig:position2}). It is noticeable that AxDNNs are more sensitive to faults than AccDNNs. This is because AxDNNs inherit approximation errors from inexact computations, and the presence of a fault in those computations exacerbates the accuracy loss (see Section \ref{subsubsec:approximationdegree}). Also, a fault in the MSB causes a drastic change in the multiplier's output value, whereas a fault in the LSB increases or decreases the value by a small amount, i.e., [0 1]. That is why the faults in MSBs lead to significant accuracy loss compared to faults in LSBs.

\subsubsection{Impact of the fault type}
We also identify the type of stuck-at faults that changes the approximation values in AxDNNs significantly. The comparison of Fig. \ref{fig:position} and Fig. \ref{fig:position2} reveals that stuck-at 1 faults are more perturbing than stuck-at 0 faults. For example, a stuck-at 1 fault in MSB causes up to 98\% - 100\% accuracy loss but a stuck-at 0 fault in MSB causes slightly lower up to 40\% accuracy loss in approximate MP-tanh (see Fig. \ref{subfig:position_a}). Similarly, a stuck-at 0 fault in MSB causes a slightly lower up to 50\% accuracy loss in approximate CC-Alx (see Fig. \ref{subfig:position_2b}). Similar trends are observed with FP-tanh (see Fig. \ref{subfig:position_2c}) and CC-VGG (see Fig. \ref{subfig:position_2d}) classification. The reason is that a stuck-at-1 fault can significantly increase the multiplier's output value by changing the bits from 0 to 1, leading to an extreme neuron activation during the forward pass. On the other hand, the stuck-at-1 fault decreases the multiplier's output value by changing the bits from 1 to 0. A simple example of this interesting fact is that a stuck-at-1 fault at the first-bit position in the 4-bit output ‘0001’ of a multiplier increases the value by 8 digits (in decimal), i.e., ‘1001’. However, a stuck-at 0 fault at the fourth position reduces the value to zero. 

\begin{figure*}[!t]
     \centering
     \vspace{-0.05in}
     \begin{subfigure}[b]{0.32\textwidth}
         \centering
         \includegraphics[width=1\textwidth]{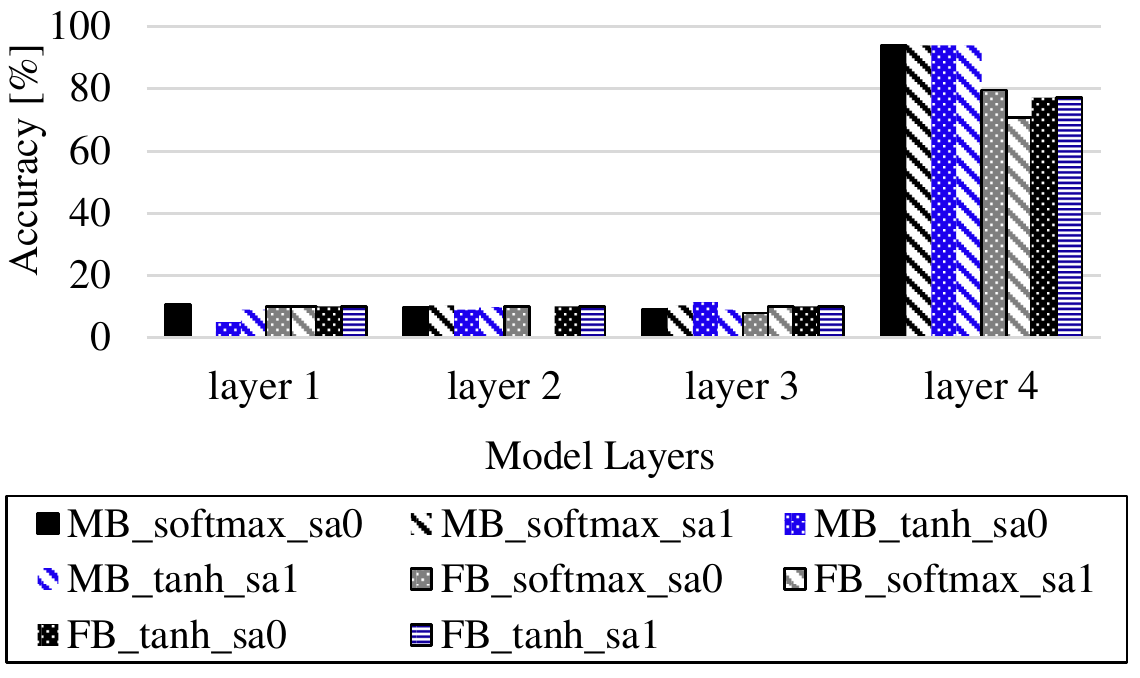}
         \caption{M1-based AxDNN Classification}
         \label{subfig:layer_a}
     \end{subfigure}
     \hfill
     \begin{subfigure}[b]{0.32\textwidth}
         \centering
         \includegraphics[width=1\textwidth]{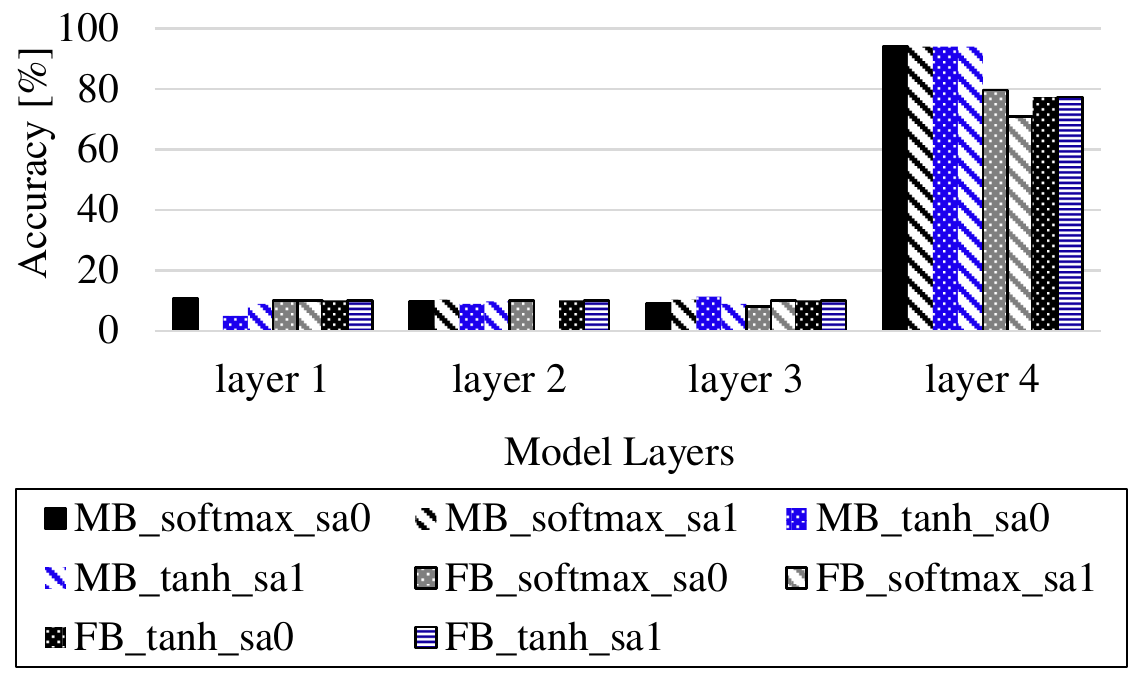}
         \caption{M2-based AxDNN Classification}
         \label{subfig:layer_b}
     \end{subfigure}     
     \hfill
     \begin{subfigure}[b]{0.32\textwidth}
         \centering
         \includegraphics[width=1\textwidth]{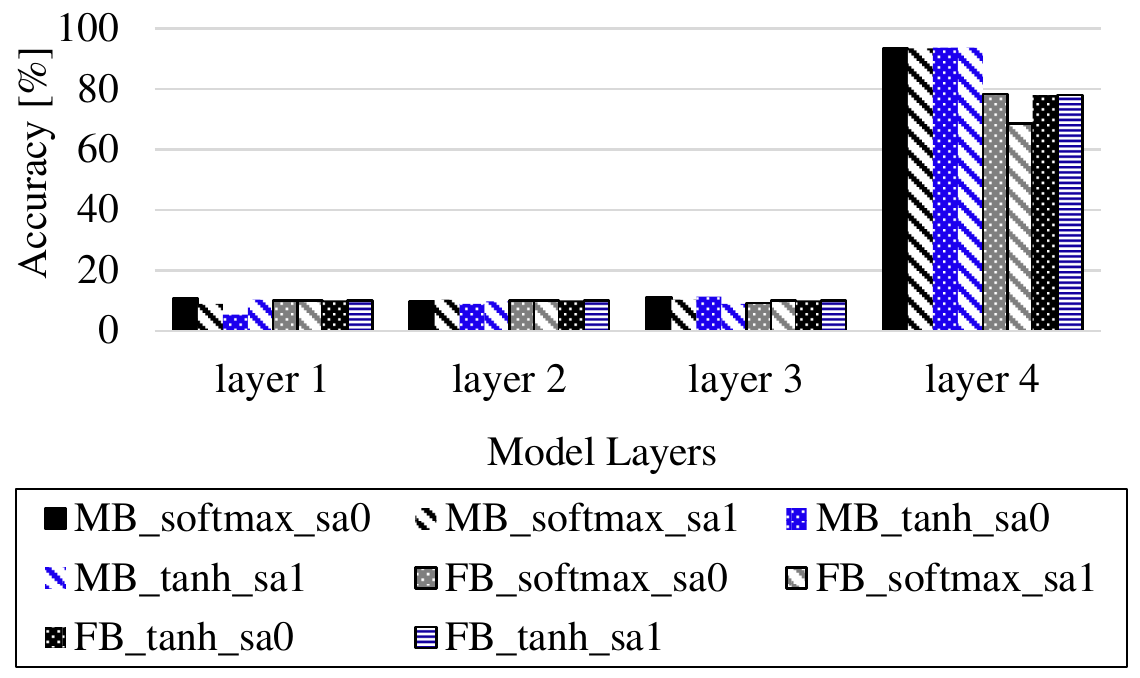}
         \caption{M3-based AxDNN Classification}
         \label{subfig:layer_c}
     \end{subfigure}     
    \vspace{-0.1in}
\caption{Impact of stuck-at 1 (sa1) and stuck-at 0 (sa0) faults on approximate multipliers M1 (0.0 \% MAE), M2 (0.0064 \% MAE) and M3 based (0.019 \% MAE) MP-tanh  and FP-tanh classification when they are injected in different layers. The MAE of each multiplier in AxDNNs is written at the bottom.}
\label{fig:layer}
\end{figure*}   

\begin{figure*}[!t]
     \centering
     \vspace{-0.01in}
     \begin{subfigure}[b]{0.32\textwidth}
         \centering
         \includegraphics[width=1\textwidth]{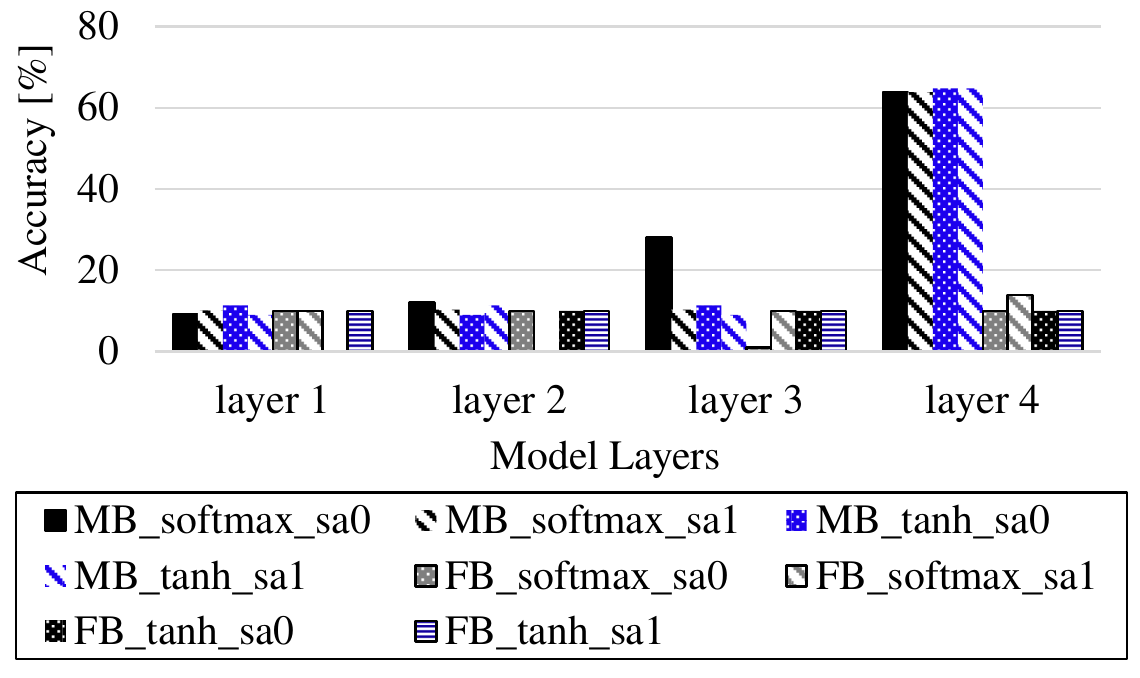}
         \caption{M4-based AxDNN Classification}
         \label{subfig:layer2_d}
     \end{subfigure}     
     \hfill
     \begin{subfigure}[b]{0.32\textwidth}
         \centering
         \includegraphics[width=1\textwidth]{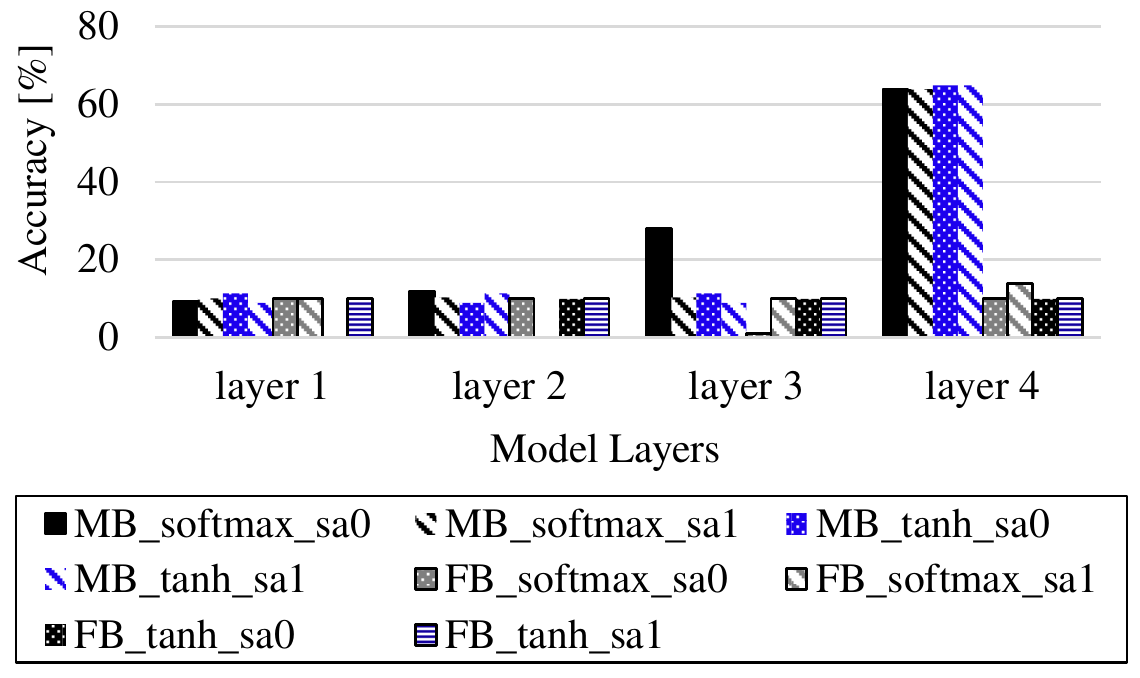}
         \caption{M5-based AxDNN Classification}
         \label{subfig:layer2_e}
     \end{subfigure}     
     \hfill
     \begin{subfigure}[b]{0.32\textwidth}
         \centering
         \includegraphics[width=1\textwidth]{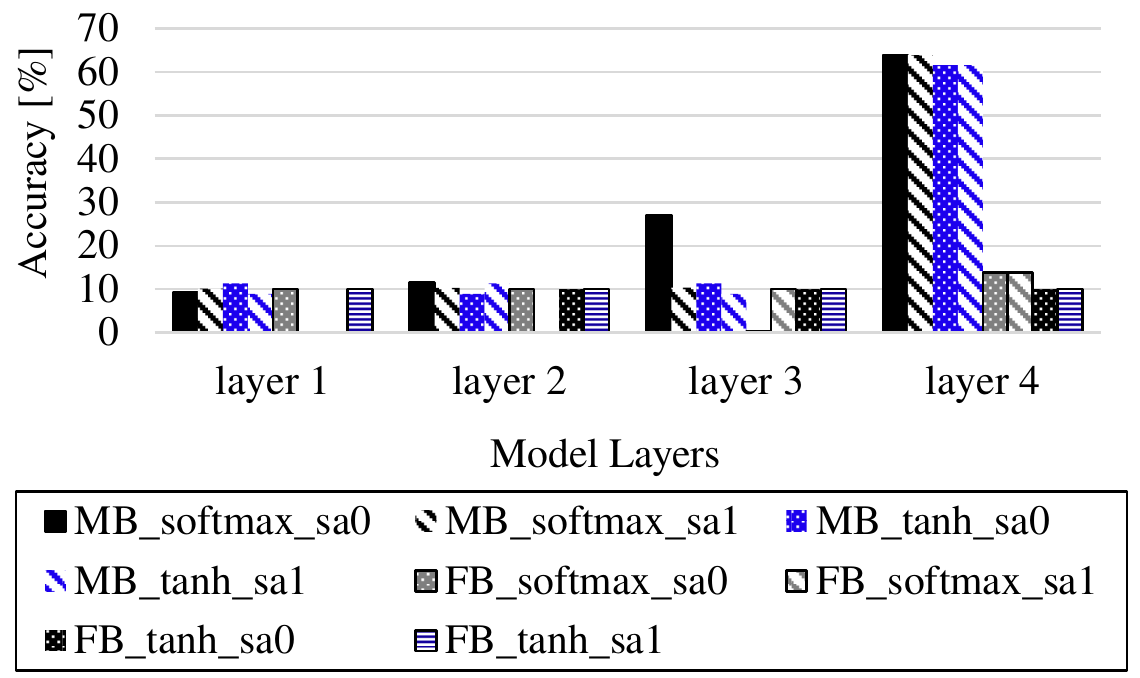}
         \caption{M6-based AxDNN Classification}
         \label{subfig:layer2_f}
     \end{subfigure}     
    \vspace{-0.01in}
\caption{Impact of stuck-at 1 (sa1) and stuck-at 0 (sa0) faults on approximate multipliers  M4 (0.23 \% MAE), M5 (0.34 \% MAE) and M6 based (0.52 \% MAE) based MP-tanh and FP-tanh classification when they are injected in different layers. The MAE of each multiplier in AxDNNs is written at the bottom.}
\label{fig:layer_2}
\end{figure*}    

\begin{figure*}[!t]
     \centering
     \vspace{-0.01in}
     \begin{subfigure}[b]{0.32\textwidth}
         \centering
         \includegraphics[width=1\textwidth]{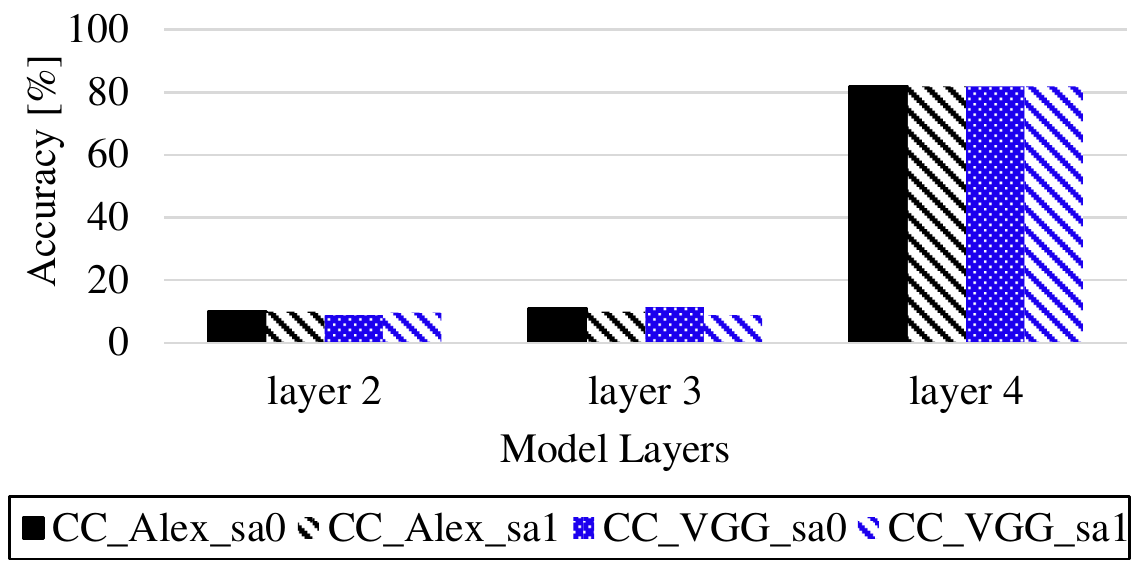}
         \caption{M1 based AxDNN Classification}
         \label{subfig:layer3_a}
     \end{subfigure}
     \hfill
     \begin{subfigure}[b]{0.32\textwidth}
         \centering
         \includegraphics[width=1\textwidth]{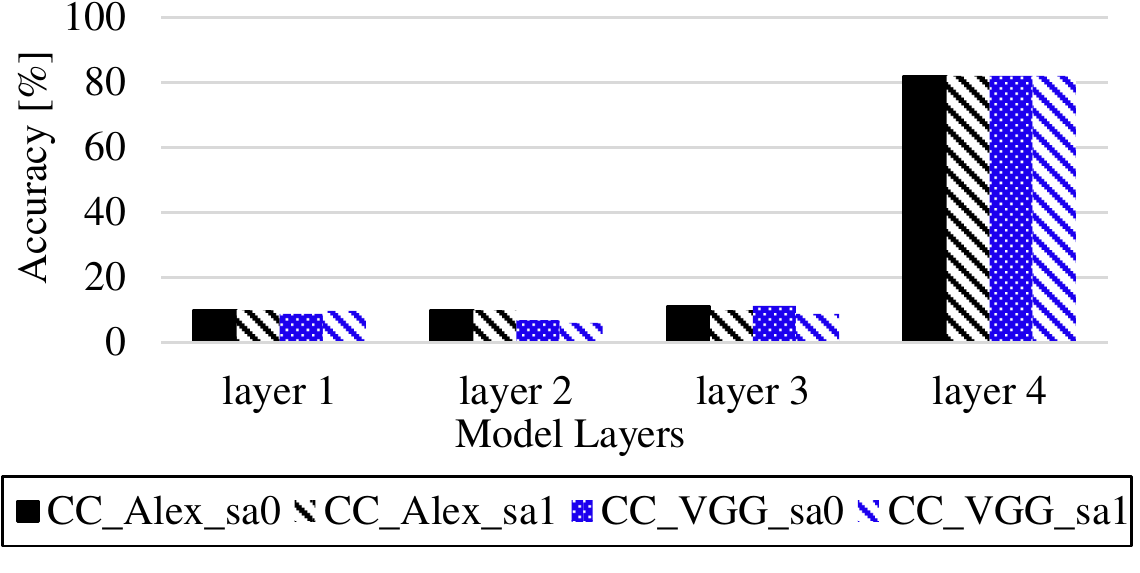}
         \caption{M2-based AxDNN Classification}
         \label{subfig:layer3_b}
     \end{subfigure}     
     \hfill
     \begin{subfigure}[b]{0.32\textwidth}
         \centering
         \includegraphics[width=1\textwidth]{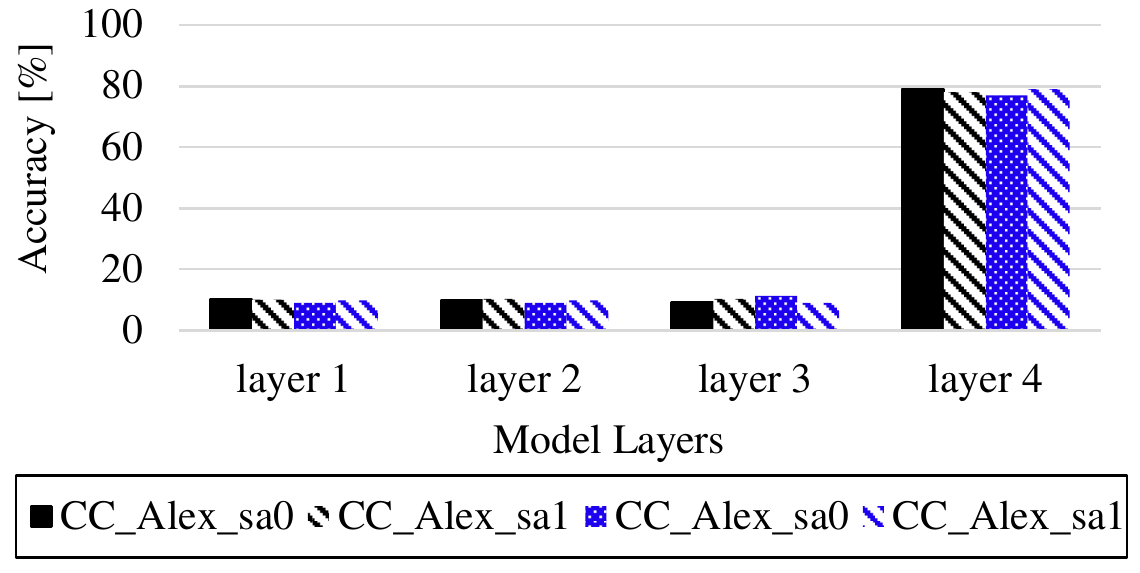}
         \caption{M3-based AxDNN Classification}
         \label{subfig:layer3_c}
     \end{subfigure}     
    \vspace{-0.05in}
\caption{Impact of stuck-at-1 (sa1) and stuck-at-0 (sa0) faults on approximate multipliers  M1 (0.0 \% MAE), M2 (0.0064 \% MAE) and M3 based (0.019 \% MAE) based CC-Alex and CC-VGG classification when they are injected in different layers. The MAE of each multiplier in AxDNNs is written at the bottom.}
\label{fig:layer_3}
\vspace{-0.01in}
\end{figure*}    

\begin{figure*}[!t]
     \centering
     \begin{subfigure}[b]{0.32\textwidth}
         \centering
         \includegraphics[width=1\textwidth]{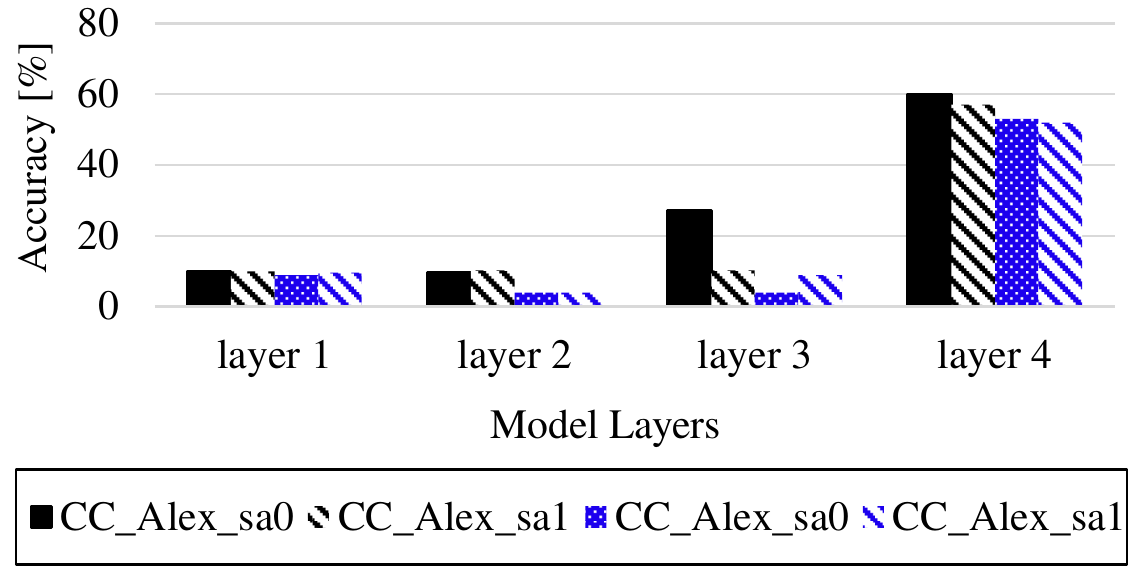}
         \caption{M4-based AxDNN Classification}
         \label{subfig:layer3_d}
     \end{subfigure}     
     \hfill
     \begin{subfigure}[b]{0.32\textwidth}
         \centering
         \includegraphics[width=1\textwidth]{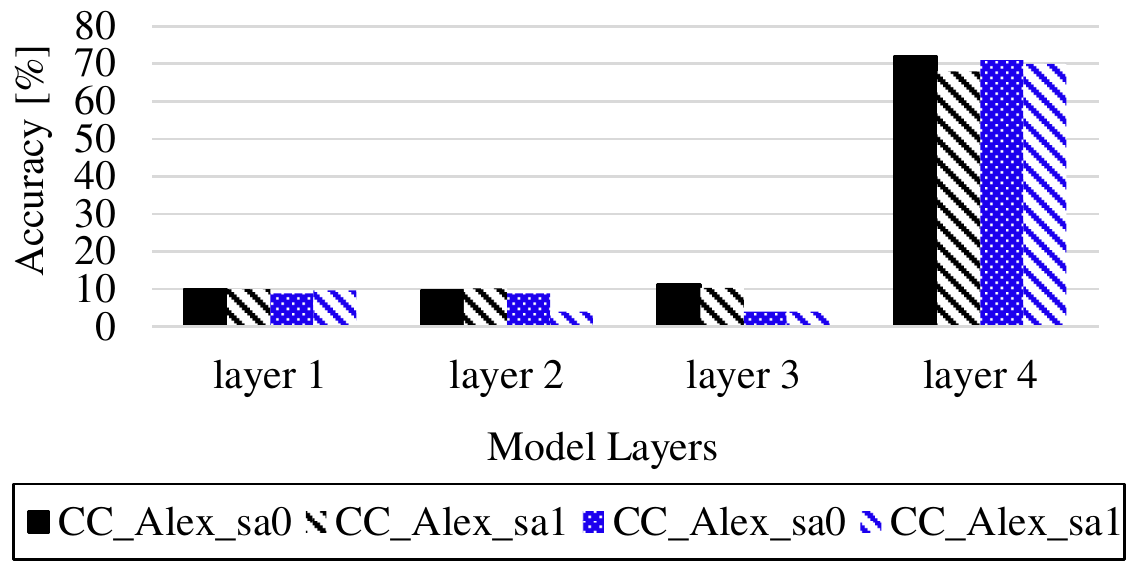}
         \caption{M5-based AxDNN Classification}
         \label{subfig:layer3_e}
     \end{subfigure}     
     \hfill
     \begin{subfigure}[b]{0.32\textwidth}
         \centering
         \includegraphics[width=1\textwidth]{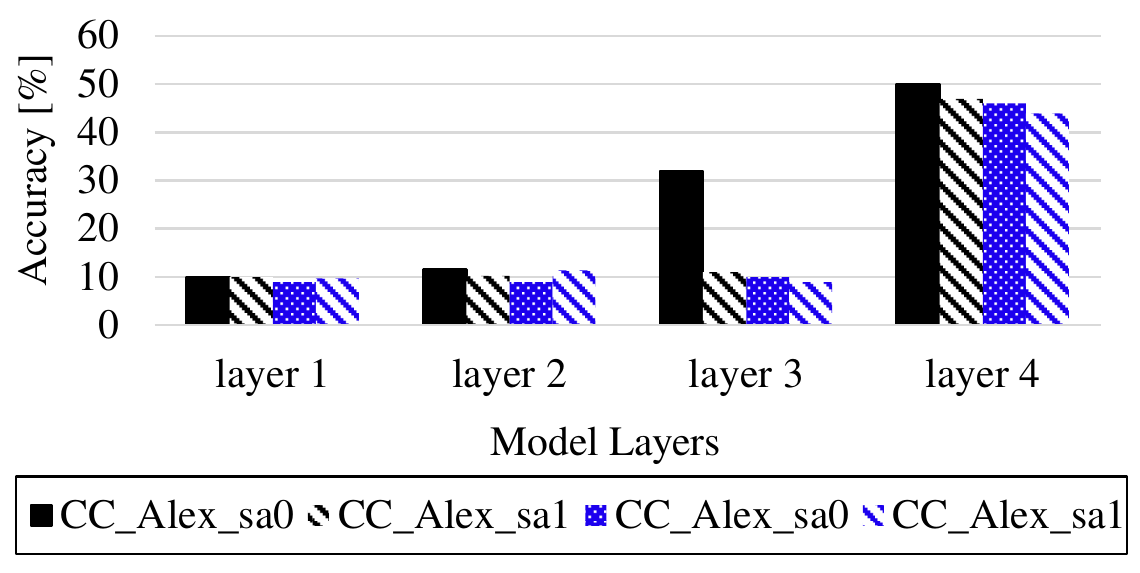}
         \caption{M6-based AxDNN Classification}
         \label{subfig:layer3_f}
     \end{subfigure}     
    \vspace{-0.01in}
\caption{Impact of stuck-at 1 (sa1) and stuck-at 0 (sa0) faults on approximate multipliers M4, M5 and M6 based CC-Alex and CC-VGG classification when they are injected in different layers. The MAE of each multiplier in AxDNNs is written at the bottom.}
\label{fig:layer2_3}
\end{figure*}    

 \begin{figure}[!t]
 	\centering
 	\vspace{-0.2in}
 	\includegraphics[width=0.9\linewidth]{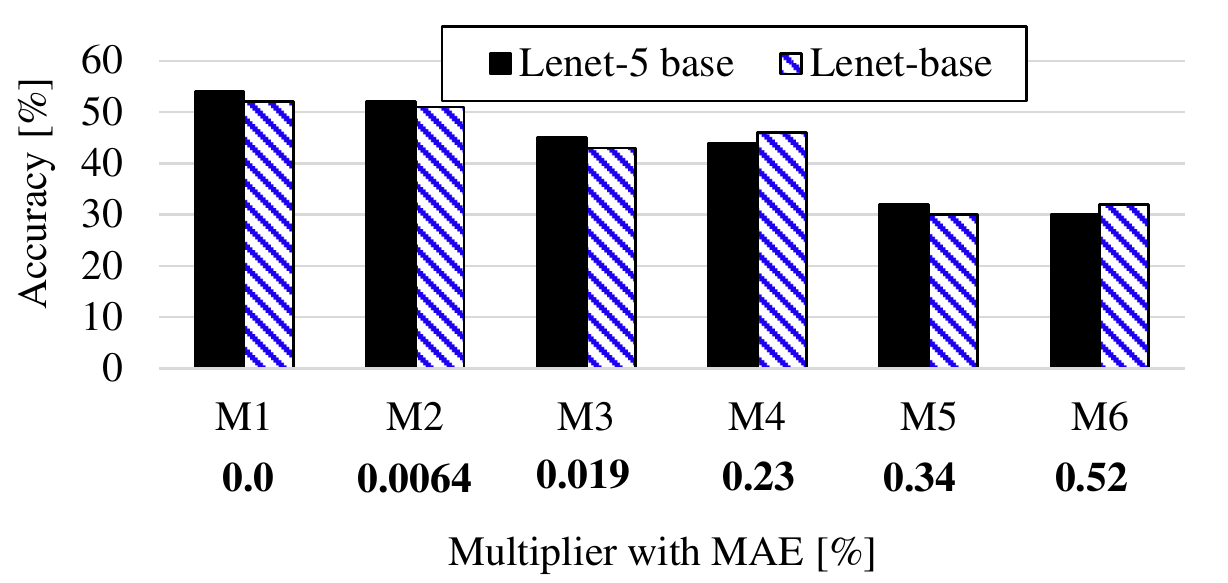}
 	\caption{Impact of stuck-at 1 (sa1) and stuck-at 0 (sa0) faults on approximate multipliers M\textit{n} based AxDNNs with different layer width. The MAE of each multiplier in AxDNNs is written at the bottom.}
 	\label{fig:wide}
 	\vspace {-0.2in}
 \end{figure}

\subsection{Fault Analysis with DNN specifications}
\label{subsec:bitwise}
In this section, we investigate how various properties of AxDNNs models affect their variability to stuck-at faults.

\subsubsection{Impact of the fault layer}
\label{subsubsec:faultRes}
We performed layer-wise fault injection analysis to identify the most and least resilient layers contributing to the overall performance efficiency of AxDNNs. This effect is more visible with stuck-at faults in MSBs. That is why this analysis is considered in this paper. Our results in Fig. \ref{fig:layer} - \ref{fig:layer2_3} illustrate that input layer 1 is relatively less resilient to the faults than the other layers, and the output layer is comparatively more fault resilient. The reason is that faults in the input layer may affect the output of all DNN layers and decrease the fault resilience significantly. For example, the stuck-at 1 faults in M6 approximate multiplier in layer 1 leads to approximately 82\% and 56\% accuracy loss, but in layer 4 approximately 30\% and 3\% accuracy loss only in accurate and approximate MP-tanh, respectively. Likewise, the stuck-at 1 faults in M5 approximate multiplier in layers 1 and 4 leads to approximately 80\% and 66\% accuracy loss approximate FP-softmax, respectively. However, the same fault configuration in accurate MP-softmax only leads to a 9\% accuracy loss. A similar trend is observed with CC-Alex and CC-VGG classification. 

\subsubsection{Impact of the activation function}
The comparison of MP-tanh and FP-tanh in Fig. \ref{fig:layer} and Fig. \ref{fig:layer_2} shows that the impact of faults varies with the activation functions in AxDNNs. We observed that architecture with tanh activation function in its hidden layers seems to be comparatively less disturbed by the faults than the one with a softmax activation function. For example, the MP-tanh and MP-softmax classification with a stuck-at-0 fault in the M1 multiplier in layer 1 results in 8.27\% and 15.14\% accuracy, respectively. Likewise, the same fault configuration for the M3 multiplier yields 9.12\% and 16.41\% accuracy, respectively. The same trend is observed in the case of the FP-tanh classification.

\subsubsection{Impact of the layer width}
We analyzed the impact of widening AxDNNs on their fault resilience by changing their layer sizes. All the convolutional and fully-connected layers in the wide models are twice as wide as the corresponding layer in the base model. In this analysis, we compare approximate ML-base and ML-wide models because increasing the layer widths slows down the AxDNN simulations. We injected the stuck-at faults in the MSB of different faulty MAC units (16\% of the total number of MAC units in systolic array) in these models. In Fig. \ref{fig:wide}, our results show that widening AxDNNs does not change their fault resilience considerably, which indicates that the approximate ML-base performs very close to approximate ML-wide even in the presence of faults. The accurate versions of both models are trained with 97\% baseline accuracy, and the accuracy drops to 52\% - 54\% with fault injection. Interestingly, the difference between their classification accuracy remains the same even with the approximation error. That means both slim and wide models have almost the same fault resilience, and the slight difference is independent of approximation error. This is because the number of vulnerable parameters grows proportionally with model width, and, as a result, the fault resilience remains constant.

\begin{figure}[!t]
	\centering
	\vspace{-0.15in}
	\includegraphics[width=1\linewidth]{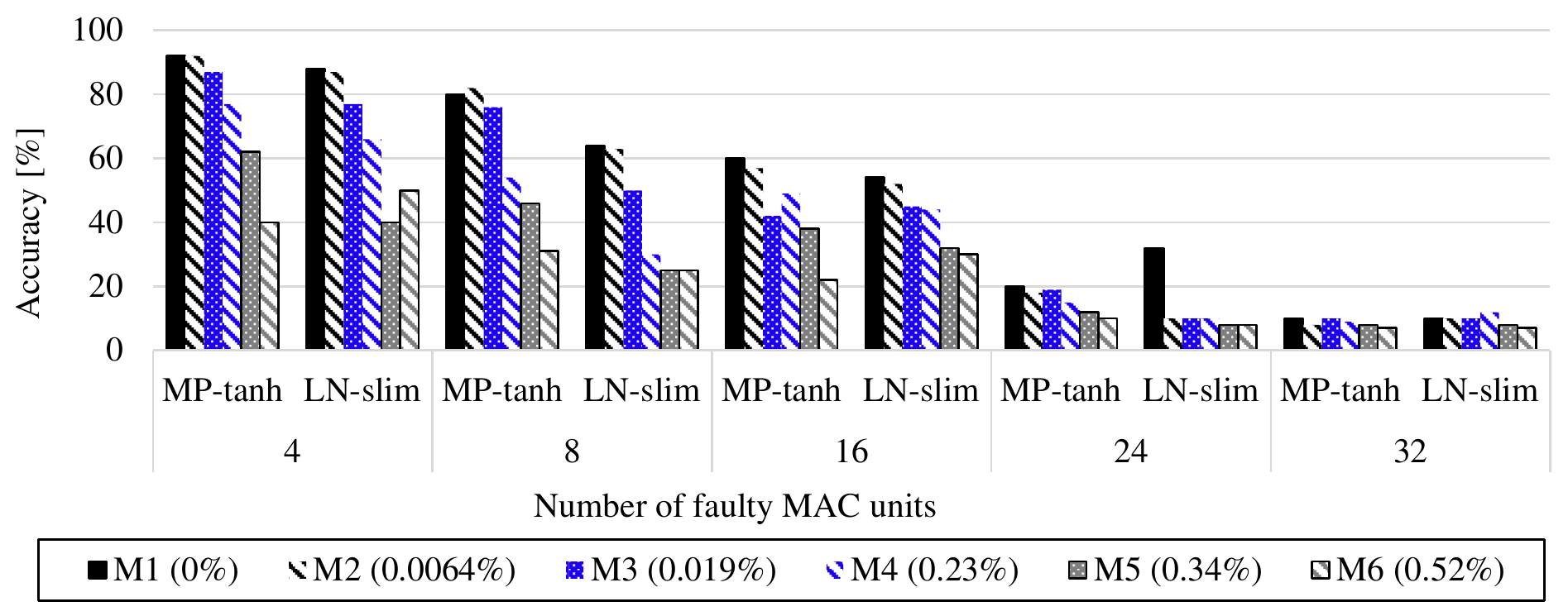}
	\caption{Impact of stuck-at faults on approximate multipliers M\textit{n} based approximate MLP-tanh vs. approximate ML-base when they are injected in 16\% MAC units in TPU. The MAE of each multiplier is written next to it.}
	\label{fig:mlpvscnn}
	\vspace{-0.15in}
\end{figure}

\subsubsection{Impact of the model architecture}
Fig. \ref {fig:mlpvscnn} compares MNIST classification with approximate MP-tanh and ML-base models with stuck-at faults injected in different percentage number of MAC units in TPU. Our results demonstrate that approximate CNNs are less resilient to stuck-at faults than approximate MLPs. For example, the classification accuracy of approximate MP-tanh is up to 26\% lower than that of approximate ML-base, especially when 8\% MAC units are faulty.

\begin{figure*}[!t]
	\centering
	\vspace{-0.15in}
	\includegraphics[width=1\linewidth]{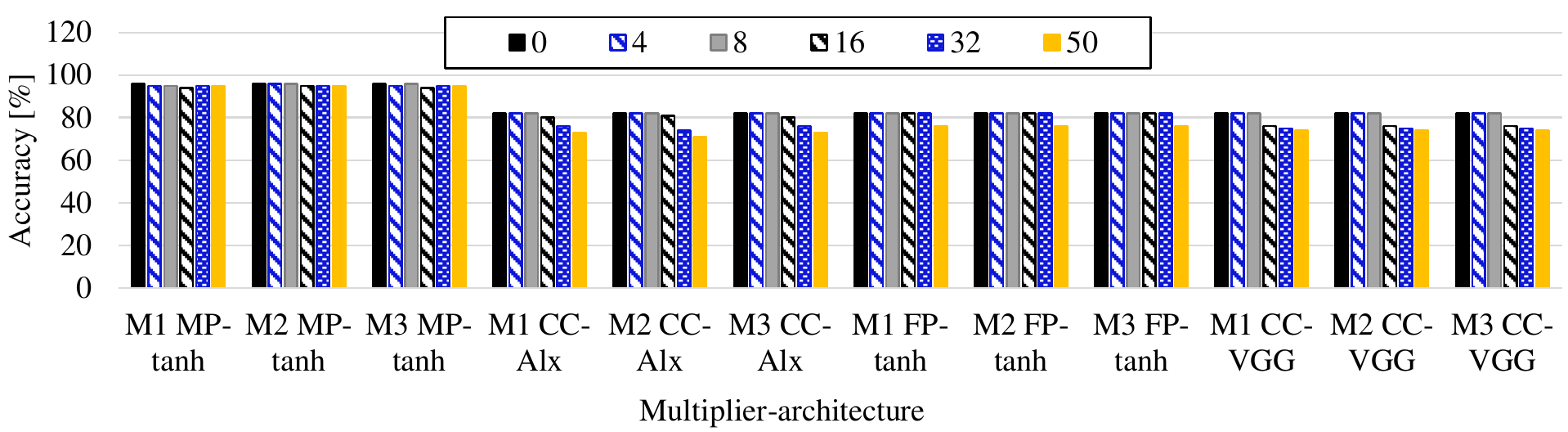}
	\caption{Fal-reTune mitigation with the stuck-at faults in approximate multipliers M\textit{n} based approximate MP-tanh, CC-Alx, FP-tanh and CC-VGG. The faults are injected in 0\%, 4\%, 8\%, 16\% and 32\% MAC units of the systolic array.}
	\label{fig:mitigation}
	\vspace{-0.1in}
\end{figure*}	

\subsection{Fault Mitigation Analysis}
To analyze the effectiveness of our proposed Fal-reTune mitigation method, we considered the case when the stuck-at faults are injected in M1-based AxDNN, M2-based AxDNN and M3-based AxDNN with different percentage number of MAC units in the systolic array. Note, the results for M4, M5 and M6 are not included as their accuracy is too low before fault injection in CC-Alx, FP-tanh and CC-VGG classification in Section \ref{subsubsec:acceleratortype}. It is worth recalling from Section \ref{subsubsec:acceleratortype} that even 32\% faulty MAC units lead to 8\%. Interestingly, Fal-reTune recovers the accuracy close to the baseline accuracy of AccDNN in AxDNN even with up to 50\% faulty MAC units as shown in Fig. \ref{fig:mitigation}. Interestingly, the recovered accuracy in the case of AxDNNs is almost the same as in the case of AccDNNs due to the weight mapping strategy for approximate multipliers that nullifies the approximation error in the Fal-reTune algorithm. For example, the accuracy remains as high as 98\%, 82\%, 81 and 82\% in case of MP-tanh, CC-Alx, FP-tanh and CC-VGG, respectively, regardless the type of the multiplier (accurate or approximate) used. We also observe that sometimes even after fault mitigation, the accuracy decreases with the increase in the percentage number of faulty MAC units. The reason is that an excessive number of MAC units can get disconnected and hence, it becomes difficult to achieve a better classification accuracy again with retraining. For example, the accuracy with 50\% faulty MAC units remains 1-6\% low when compared to the mitigation is applied to AxDNNs mapped to a systolic array with 4\% faulty MAC units.

\subsection{Fault-Energy Tradeoffs Exploration}
\begin{figure}[!t]
 	\centering
 	\vspace{-0.05in}
 	\includegraphics[width=0.8\linewidth]{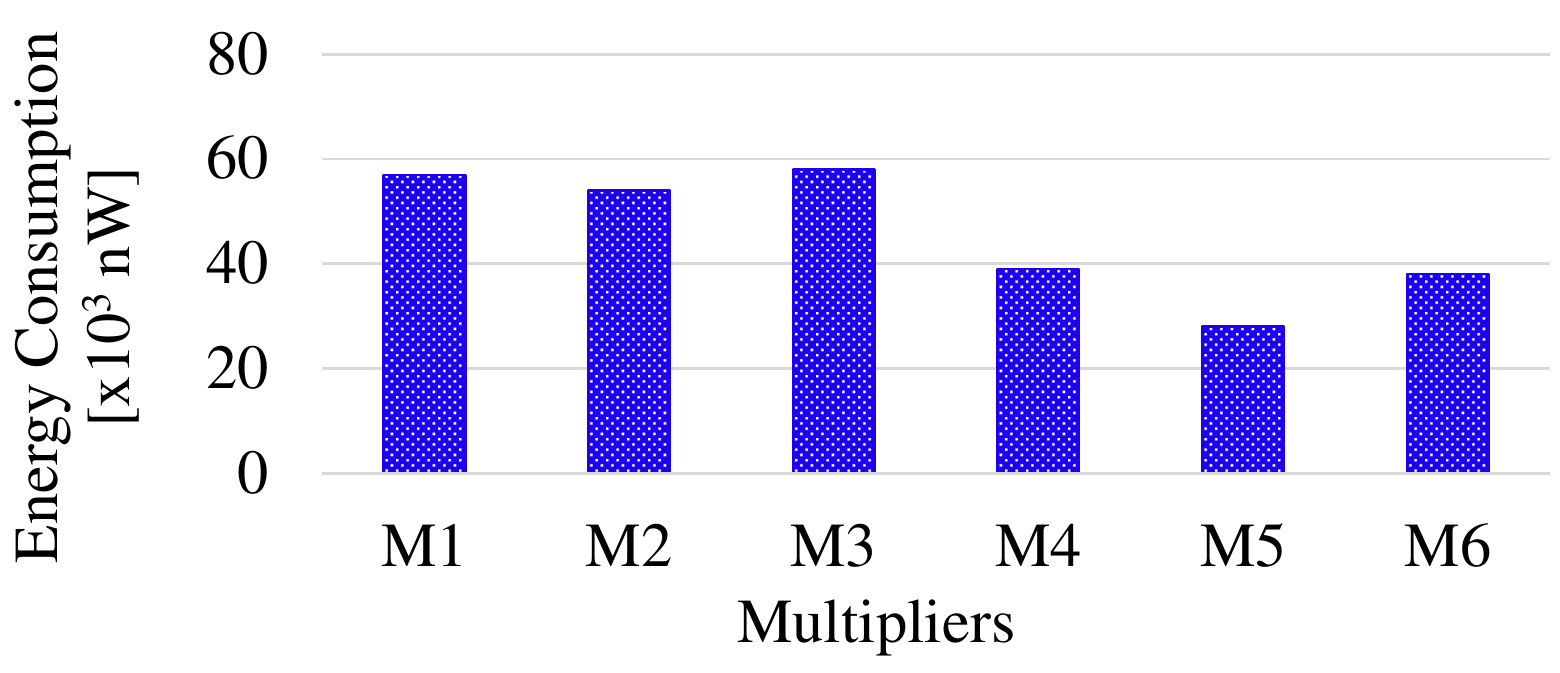}
 	\caption{Energy Analysis of 8x8 approximate systolic arrays M1 to M6 multipliers from Evoapprox8b \cite{mrazek2017evoapproxsb} library}
 	\label{fig:energy}
 	\vspace {-0.15in}
 \end{figure}

Since simulating a 256x256 systolic array requires lots of hardware resources; therefore, we analyze the energy consumption of an 8x8 systolic array using M1 to M9 multipliers from Evoapprox8b \cite{mrazek2017evoapprox8b} library. Our results in Fig. \ref{fig:energy} show that M2-based AxDNNs are less energy efficient and M4, M5, and M6-based AxDNNs are the most energy efficient among all AxDNNs studied in this paper. On the contrary, it is obvious from the discussion in Section \ref{subsesc:acceleratorspec} that M2-based AxDNNs are the most fault resilient and M4, M5, and M6-based AxDNNs are the least fault resilient among all AxDNNs studied in this paper. \textit{Hence, the fault resilience and energy efficiency are orthogonal to each other.} Furthermore, a slight difference in the energy consumption of M1 and M2-based AxDNNs is observed. Though, M6-based AxDNN is less fault resilient than M1-based AxDNN. M6-based AxDNN is observed quite an energy efficient and fault resilient as it exhibits above 90\% and 80\% accuracy, with faulty LSB in the input layer, in the MNIST \cite{deng2012mnist} and Fashion MNIST \cite{xiao2017fashion} classification (as shown in Fig. \ref{fig:layer} and Fig. \ref{fig:layer_2}), respectively. \\

\section{Conclusion}
\label{sec:conclusion}

Approximate computing trades the classification accuracy of AxDNNs with energy efficiency for performance gains in error-resilient applications. However, the inexactness caused by the approximation errors in AxDNNs scales down their fault resilience. This paper extensively analyzes and mitigates the reliability degradation due to permanent faults in approximate feed-forward neural networks such as MLP and CNN using the state-of-the-art Evoapprox8b \cite{mrazek2017evoapproxsb} library. We propose a novel fault mitigation method i.e., fault-aware retuning of weights (Fal-reTune) that retunes the weights using a weight mapping function in the presence of faults for improved classification accuracy. To evaluate the fault resilience and the effectiveness of our proposed mitigation method, we used the most widely used MNIST, Fashion-MNIST, and CIFAR10 datasets. Our results show that the faults affect the classification accuracy of AxDNNs more than AccDNNs. For instance, a permanent fault in AxDNNs can lead up to 56\% accuracy loss, whereas the same faulty bit can lead to only 4\% accuracy loss in AccDNN. Our proposed Fal-reTune mitigation efficiently improves the classification accuracy of up to 98\% even with fault rates of up to 50\%.


\bibliographystyle{IEEEtran}
\bibliography{bib/conf}

\begin{IEEEbiography}[{\includegraphics[width=1in,height=1.25in,clip,keepaspectratio]{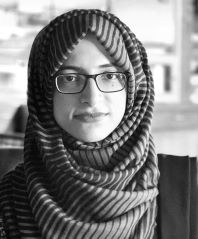}}]{Ayesha Siddique} received her B.Sc. degree in electrical engineering from the Army Public College of Management and Sciences, Rawalpindi, Pakistan, in 2013, and her M.S. degree in electrical engineering from the National University of Sciences and Technology (NUST), Islamabad, Pakistan, in 2018. She is currently pursuing her Ph.D. degree in electrical and computer engineering and working as a graduate research assistant at the Dependable Cyber-Physical Systems (DCPS) Laboratory in the University of Missouri-Columbia. She also served as a Research Associate and Design Engineer with the Center for Advanced Research in Engineering, Islamabad, Pakistan, from 2013 to 2017 after her Bachelor's degree. She is a recipient of the Richard Newton Fellowship at DAC 2018. Her research interests include formal analysis and verification of embedded systems, reliable and secure machine learning, energy-efficient systems, and digital image processing.
\end{IEEEbiography}

\vskip -1\baselineskip plus -1fil

\begin{IEEEbiography}[{\includegraphics[width=1in,height=1.25in,clip,keepaspectratio]{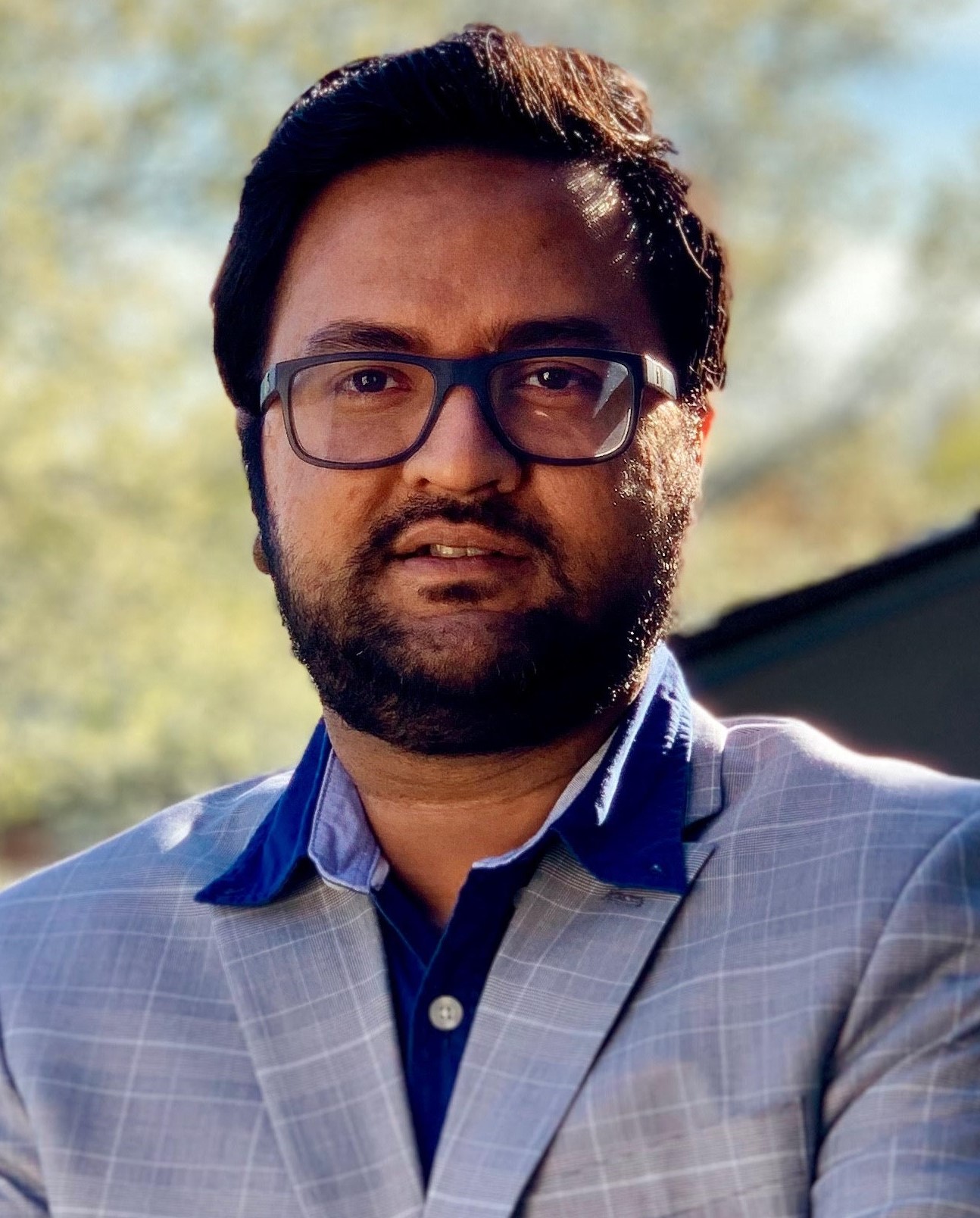}}]{Khaza Anuarul Hoque} received his M.Sc. and Ph.D. degrees from the Department of Electrical and Computer Engineering at Concordia University, Montreal, Canada in 2011 and 2016, respectively. He is currently an Assistant Professor in the Department of Electrical Engineering and Computer Science at the University of Missouri-Columbia (MU) where he directs the Dependable Cyber-Physical Systems (DCPS) Laboratory. His research interests include Cyber-physical and Embedded Systems, Machine Learning, Cybersecurity, and Formal Methods. Before joining MU, he was an FRQNT postdoctoral fellow at the University of Oxford, UK. Dr. Hoque has received several awards and distinctions, including FQRNT Postdoctoral Fellowship Award (2016), FQRNT Doctoral Research Award (2012), and Best Paper Award in 8th IEEE International NEWCAS Conference (2011). He is a senior member of IEEE, and a member of ACM, AAAS, and IEEE Technical Committee on Cyber-Physical Systems (CPS). 
\end{IEEEbiography}

\end{document}